# 6G Resilience

WHITE PAPER

EDITORS: HIRLEY ALVES, NURUL H. MAHMOOD, ONEL L. A. LÓPEZ, SUMUDU SAMARAKOON, SEPPO YRJÖLÄ, MATTI LATVA-AHO, MARKKU JUNTTI, ARI POUTTU

# Executive summary

6G must be designed to withstand, adapt to, and evolve amid prolonged, complex disruptions. Mobile networks' shift from efficiency-first to sustainability-aware has motivated this white paper to assert that **resilience** is a primary design goal, alongside sustainability and efficiency, encompassing technology, architecture, and economics.

We promote resilience by analysing dependencies between mobile networks and other critical systems, such as energy, transport, and emergency services, and illustrate how cascading failures spread through infrastructures. We formalise resilience using the 3R framework: *reliability, robustness, resilience*. Subsequently, we translate this into measurable capabilities: graceful degradation, situational awareness, rapid reconfiguration, and learning-driven improvement and recovery.

Architecturally, we promote edge-native and locality-aware designs, open interfaces, and programmability to enable islanded operations, fallback modes, and multi-layer diversity (radio, compute, energy, timing). Key enablers include AI-native control loops with verifiable behaviour, zero-trust security rooted in hardware and supply-chain integrity, and networking techniques that prioritise critical traffic, time-sensitive flows, and inter-domain coordination.

Resilience also has a techno-economic aspect: open platforms and high-quality complementors generate ecosystem externalities that enhance resilience while opening new markets. We identify nine business-model groups and several patterns aligned with the 3R objectives, and we outline governance and standardisation.

This white paper serves as an initial step and catalyst for 6G resilience. It aims to inspire researchers, professionals, government officials, and the public, providing them with the essential components to understand and shape the development of 6G resilience.

## Editors

Hirley Alves, Nurul H. Mahmood, Onel L. A. López, Sumudu Samarakoon, Seppo Yrjölä, Matti Latva-Aho, Markku Juntti, and Ari Pouttu

## Contributors

| | | |
|---|---|---|
| Armin Dekorsy | Harri Saarnisaari | Nhan Nguyen |
| Arthur Sousa de Sena | Hirley Alves | Nurul H. Mahmood |
| Aydin Sezgin | Italo Atzeni | Onel L. A. López |
| Bho Matthiesen | Jaap Van De Beek | Pawani Porambage |
| Chafika Benzaid | Jacek Rak | Petar Popovski |
| Chathuranga Weeraddana | Konstantin Mikhaylov | Petri Ahokangas |
| David Hutchison | Lauri Loven | Premanandana Rajatheva |
| Dileepa Marasinghe | Madhusanka Liyanage | Robert-Jeron Reifert |
| Doganalp Ergenc | Marcos Katz | Seppo Yrjölä |
| Eduard Jorswieck | Marja Matinmikko-Blue | Sumudu Samarakoon |
| Erkki Harjula | Matti Latva-Aho | Tharaka Hewa |
| Falko Dressler | Mehdi Rasti | Tommy Svensson |
| | Mika Ylianttila | |

# Table of Contents

# 1  Motivation

Mobile connectivity is a critical infrastructure of the modern world. In 2024, approximately 5.8 billion people, 71% of the global population, subscribed to mobile services, with 4.7 billion accessing the internet via mobile networks. The mobile communications sector contributes 5.8% of global GDP, equivalent to €6.5 trillion added in 2024, underscoring its significant role in shaping economies, facilitating digital participation, and connecting communities across continents. The society is becoming increasingly digitised and dependent on real-time wireless communication, such that the resilience of these networks emerges as a critical requirement.

> **Resilience:** "… ability to recover, adapt, and evolve in the face of challenges." See Chapter 3.

The expanding reliance on wireless connectivity brings new vulnerabilities, as localised failures can escalate into systemic disruptions of essential services, including healthcare, emergency response, and transportation. Resilience is therefore a non-optional design objective: networks must sustain essential functions, degrade gracefully, and recover rapidly. Traditional priorities such as speed, latency, and capacity must now be balanced with the ability to provide robust, secure, and uninterrupted service under stress.

While past generations of mobile networks have mainly focused on enhancing performance and efficiency, often through computationally intensive optimisations, the concept of resilience has remained underexplored. Disruptions from cyberattacks, natural disasters, system overloads, and unforeseen faults can lead to cascading failures with significant societal and economic consequences. As the world prepares for the next leap in wireless technology, resilience must be elevated to a first-order design principle.

## 1.1 A new paradigm for 6G

The upcoming 6$^{th}$ generation of mobile networks (6G) presents an opportunity to fundamentally reimagine how wireless systems are designed, deployed, and managed. Resilience is not an emergent by-product, but a first-class objective supported by capabilities to anticipate, absorb, adapt, and recover, while meeting sustainability goals. Achieving this vision requires the convergence of multiple disciplines and technologies.

Four broad research priorities are emerging:

- **Resilient-by-design principles**
  Developing adaptive network architectures that can anticipate and withstand failures, using Artificial Intelligence (AI)-driven anomaly detection, self-organising topologies, and redundancy-aware protocols.
- **AI-optimised radio access networks (AI-RAN)**

Leveraging AI for predictive maintenance, dynamic spectrum use, autonomous fault recovery, and real-time reconfiguration to ensure service continuity.
- **Energy Efficiency and Sustainable Coverage**
Designing energy-aware algorithms and integrating renewable-powered infrastructure, including terrestrial and non-terrestrial elements, to enhance both environmental sustainability and operational robustness.
- **Interplay of Resilience, Efficiency, and Sustainability**
Understanding and navigating the trade-offs among these three cornerstones is crucial for designing resilient, high-performing, sustainable and economically viable networks.

There are many research challenges ahead; current research lacks i) a compact, comparable metric set for service continuity and recovery (e.g., detection and repair times, graceful-degradation targets), ii) validated test methods spanning integrated terrestrial and non-Terrestrial Networks (NTNs), iii) energy-aware resilience operations, and iv) end-to-end threat-to-control-loop mappings. This white paper addresses these issues by proposing principles, metrics, and validation methods, as well as surveying enabling technologies.

The challenges are not theoretical. The future wireless ecosystem will support mission-critical services where system failure could have life-threatening or economically devastating consequences. Resilience, therefore, is not a luxury. It is a necessity.

## 1.2 Responding to global challenges

The increasing complexity of communication systems coincides with rising geopolitical tensions, climate-driven natural disasters, and growing cyber threats. Future wireless infrastructure must operate across hybrid environments, comprising terrestrial, aerial, and space-based, and remain functional even during partial outages, physical damage, or hostile interference. As energy demands rise and connectivity expands into previously underserved regions, there is a growing imperative to design solutions that are both energy-efficient and resilient.

Recent global frameworks and policy statements have emphasised the importance of secure, open, and resilient communications as crucial to economic stability and national security. Emerging global initiatives have underscored the economic value of resilient infrastructure, with studies showing a return of up to €4 for every €1 invested. However, current standardisation processes, such as those under 3GPP, have yet to incorporate resilience as a core requirement.

## 1.3 The research frontier

Despite growing recognition of its importance, resilience remains a nascent area within wireless systems research. Existing efforts have yet to comprehensively address the systemic and cross-layer nature of resilience in mobile networks. Emerging programs in North America, Europe, and Asia have begun exploring these topics, but significant gaps remain in:

- Standardising resilience metrics and evaluation frameworks.
- Integrating resilience into network slicing, edge computing, and AI-based management systems.
- Ensuring interoperability across diverse hardware, vendors, services, and platforms.
- Balancing value contributions and value appropriation sharing among stakeholders, including the platform owners, complementors, and side actors.
- Balancing resilience, efficiency, sustainability and cost constraints.
- Identifying and understanding the dependencies and vulnerabilities in interconnected systems.

To build truly resilient 6G systems, the global research community must collaborate across disciplines, combining communications engineering, cybersecurity, AI, power systems, and human-centred design. Recently, the ITU-R has emphasised resilience within the IMT-2030 vision (ITU-R, 2023), and 3GPP initiated 6G system-level discussions on resilience at the Incheon workshop in March 2025 (3GPP, 2025). These indicate momentum in aligning principles, KPIs, and evaluation methods. However, a unified effort is needed to influence international standards, inform policy, and guide industry adoption.

## 1.4 From performance to resilience

The evolution from 5G to 6G marks a shift not only in speed and spectral efficiency but also in societal expectations. Mobile networks are no longer convenient layers; they are lifelines. The vision for 6G must therefore expand beyond performance, qualification, and quantification of resilience.

This white paper sets out to explore the multidimensional nature of resilience in wireless communications. It defines the technical challenges, identifies strategic research priorities, and calls for a global agenda that places resilience at the centre of 6G development. The future of connectivity will not be defined solely by how fast we can transmit data, but by how well our systems can endure, recover, and continue to serve society under pressure.

# 2 Mobile technologies as part of the critical infrastructure

Over the past few decades, mobile technologies have evolved from basic voice services to becoming the backbone of digital society. Successive generations of mobile networks, from the analogue first generation (1G) to today's advanced fifth generation (5G), have enhanced performance and capacity, unlocking transformative applications across various sectors, including public safety, healthcare, manufacturing, energy, and transportation. Mobile networks are now widely recognised as critical infrastructure, and future generations are expected to further enhance their capabilities by integrating communication with sensing and intelligence on an unprecedented scale.

As dependence on mobile connectivity deepens, the consequences of network disruptions grow more severe. Essential services increasingly rely on robust and continuous connectivity, making even brief outages a potential source of cascading economic, safety, and operational risks. Strengthening resilience in mobile networks, which is the network's ability to absorb and quickly recover from disruptions, is no longer optional—it is vital for ensuring reliable service delivery in complex and uncertain environments.

Achieving resilience requires addressing challenges across multiple layers, from the infrastructure and architecture of networks to the societal, environmental, and business ecosystems within which they operate. Achieving resilient mobile infrastructure requires addressing longstanding societal challenges, particularly the digital divide. Gaps in coverage and capacity between urban and rural regions hinder inclusion, access to essential services, and regional economic vitality. At the same time, the architectural shift toward distributed computing is reshaping how resilience is engineered. Moving processing closer to data sources through edge computing enables faster response times and greater fault tolerance. This decentralisation enhances service continuity, but it also introduces new risks related to coordination, security, and infrastructure design.

Intelligent systems are becoming central to managing the growing complexity and dynamism of mobile networks. Machine learning (ML) and AI enable predictive maintenance, autonomous fault recovery, and proactive resource management. These capabilities support a shift from reactive to adaptive and anticipatory network operations. AI also offers a path to reconcile resilience with sustainability, helping manage trade-offs such as energy consumption, redundancy, and resource allocation.

Mobile networks are also increasingly entangled with other critical infrastructures. Their reliance on satellite-based synchronisation systems, such as the Global Navigation Satellite System (GNSS), introduces vulnerabilities that can impact multiple sectors. Simultaneously, mutual dependencies between mobile networks and renewable-powered smart grids create both risks and opportunities. Cross-sector coordination is crucial to ensure the continuous operation of interdependent systems.

This chapter exposes the interwoven dimensions (that is, technological, societal, and infrastructural) that define the role of mobile networks in critical infrastructure. It sets the stage for understanding the resilience challenges and opportunities that must be addressed as we move toward the sixth generation (6G) era and beyond.

> Figure 2.1
> **Description**: info-chart describing the mobile networks' evolution from 1 to 6 G. It highlights the transition from voice to data, performance to sustainability to resilience.

*Figure 2.1 Cellular evolution - 1g to 6g – from efficiency to resilience.*

## 2.1 The digital divide

Ensuring comprehensive network coverage globally is not just a technological ambition but a fundamental necessity for societal progress and global development. The worldwide call for urban-standard broadband Internet in rural areas underscores a critical need across three vital domains. Firstly, inclusivity requires that even the most remote municipalities have access to high-speed Internet, bridging gaps in education, economic opportunity, and social inclusion. Secondly, reliable connectivity is crucial for ensuring safety and delivering healthcare services, guaranteeing continuity regardless of location. Ultimately, the economic vitality of agriculture, tourism, and energy relies on robust digital infrastructure, thereby preventing rural areas from being excluded from the digital transformation.

The widening rural-urban digital divide, exacerbated by successive generations of mobile networks, presents a persistent societal challenge. Over three decades, market-driven approaches have not adequately addressed the imperative to equalise access. Changing this status quo is essential, aligning with global goals such as the United Nations' Sustainable Development Goals (SDGs)[1]. Rural connectivity is pivotal to achieving the SDGs: from ensuring quality education (Goal 4) to promoting economic growth and industry (Goal 8), and developing sustainable cities by building healthy, interlinked rural infrastructure (Goals 11 and 3). It also facilitates climate action through innovative agriculture and transportation (Goal 13).

In many rural areas and remote regions, broadband networks are often underdeveloped. Specific challenges are:

- Ambitious national broadband strategies often set targets for universal access but lack clear implementation paths. Who ensures these targets are met?
- Emergency services are compromised when emergency calls cannot be made in areas without coverage. What guarantees mobile coverage for critical safety functions?

---

[1] https://sdgs.un.org/goals

- Agriculture and forestry depend increasingly on digital tools and IoT for automation to enhance efficiency. How can these industries remain competitive without reliable coverage in the areas where their operations are located?
- Emergency networks often require dense and costly infrastructure. Can commercial networks and public safety requirements be aligned to reduce deployment burdens while coping with more stringent quality of service (QoS) guarantees?
- Connectivity influences rural quality of life. Without digital services, people may avoid living and working in rural regions, worsening local labour shortages.
- E-health and remote care can save public funds and improve services. However, how can rural healthcare modernise without universal digital access?
- The rise of connected and autonomous transport demands full-area mobile coverage. How can regulatory and liability frameworks evolve if coverage remains fragmented?

The challenges appear in several dimensions:

- **Culture and norms.** A deep-rooted urban norm governs much of today's policies, decision-making, regulations, and solution design, limiting prioritisation of rural areas. Arguably, this is the dimension in which challenges are largest.
- **Policies and regulations**. The cellular market is strongly governed by laws and regulations, necessitated by the very nature of the radio spectrum. Hence, challenges are at the heart of the structural changes that are needed.
- **Infrastructure**. Cellular and communications networks constitute a societal infrastructure that needs to be built in rural areas. Challenges relate to clear ownership and business incentives.
- **Business models and value chains**. Value chains need to be modelled and renewed. Economic values in rural regions are enormous, but they are not adequately reflected in many of today's business models. The challenges here are to account for the potential values in rural areas that will be unleashed by proper rural Internet penetration.
- **Technology**. Arguably one of the least significant challenges, technology is available to provide full-area coverage. Technology solutions known to engineers as effective in solving challenges are not currently available on the market for reasons outlined in the above four dimensions. Piloting and showcasing effective technologies could catalyse broader change.

In summary, a wide array of societal issues, including inclusion, economic opportunity, safety, healthcare, and mobility, are closely tied to the absence of ubiquitous fixed or mobile broadband. Meeting these challenges requires coordinated action across policy, technology, and markets to ensure resilient and equitable connectivity for all.

To address these challenges, non-terrestrial networks (NTN) have emerged as a complementary solution to terrestrial infrastructure, offering promising avenues for bridging the digital divide (Alves, et al., 2024). NTNs, comprising low Earth orbit (LEO) satellite constellations, high-altitude platform stations (HAPS), and UAVs, may provide baseline coverage in underserved regions where terrestrial deployment is economically or logistically unfeasible. Notably, these systems are not primarily deployed for resilience but for coverage and sustainability. However, once in place, they inadvertently strengthen network resilience by adding additional layers of infrastructure. LEO satellites, for example, provide global coverage and essential connectivity, but they are unlikely to fulfil the increasing demand for high-capacity services on their own. Meanwhile, HAPS and UAVs offer flexible, on-demand capacity that can complement terrestrial networks without contributing to unsustainable densification. These layered approaches support key value indicators (KVIs) for 6G, that is, inclusion, sustainability, and resilience, by ensuring that all communities, regardless of geography, can participate in and benefit from digital transformation.

> Figure 2.2
> **Description**: digital divide challenges and dimensions

*Figure 2.2 Graphical illustration of challenges and dimensions of the digital divide*

## 2.2 Edge cloud continuum

The excessive centralisation of computing presents a vulnerability for digital services. In such systems, network and power outages, cyberattacks, and other disturbances can severely hinder or even prevent the use of services by reducing the quality of the connection or disconnecting users and data sources from computing locations. The role of data centres and the distributed computing continuum, specifically the edge–core–cloud continuum, is foreseen as an enabler for resilient, responsive, and scalable services. Centralised cloud architectures are increasingly complemented by edge computing and edge data centres to meet the strict latency, bandwidth, and availability requirements of foreseen 6G-enabled services, especially in critical domains such as healthcare, energy, and public safety.

Edge computing brings computation and data processing closer to the data sources and end-user devices. This way, it reduces response time and enhances resilience to, e.g., network or power outages and other unexpected disturbances, by decentralising processing and reducing dependence on core network connectivity. The increased locality ensures the continuity of critical services during disruptions. While centralised systems can be made robust against known threats, such as predictable traffic peaks or known failure modes, they are inherently more vulnerable to unforeseen disruptions, including backhaul outages, cyberattacks, or natural disasters, which can potentially prevent connectivity to centralised resources for extended periods. In contrast, when critical computing is handled closer to the point of need, such as in multi-access edge computing (MEC) nodes co-located with radio access network (RAN) base stations (edge computing nodes at the middle tier of Figure 2.3) or in power grid substations, hospitals, or emergency command centres (local edge computing clusters at the bottom tier of the Figure 3), the system becomes more resilient, better able to maintain essential operations even during unpredictable events.

For example, in the energy domain, edge computing at substations can enable autonomous voltage regulation and fault detection. Local ML-enabled controllers can identify various anomalies and isolate faults within milliseconds, preventing cascading failures across the grid, even in cases when the connectivity to central systems is disrupted. In another example, a hospital served by a private 5G network can continue its critical operations, such as triggering alerts while monitoring patient vitals, even when connectivity to the central computing infrastructure is lost, thereby maintaining critical patient care. These localised capabilities demonstrate how edge resources contribute to absorbing and recovering from both anticipated and unforeseen events, ensuring continuity and safety. Notably, these benefits can be harnessed at the cost of stable power availability at the edge; therefore, ensuring continuous energy supply is equally critical. Section **Error! Reference source not found.** discusses energy networking as an enabler for resilient edge operations.

Overall, the emerging distributed computing continuum is not just an architectural evolution aimed at improving performance; it represents a fundamental shift toward distributed resilience, which is essential for maintaining critical infrastructure operations under conditions of uncertainty.

> Figure 2.3
> **Description**: distributed computing layered architecture

*Figure 2.3 - Distributed computing continuum*

## 2.3 Intelligent systems for enhanced network resilience

As mobile networks become increasingly complex and dynamic, ensuring resilience becomes significantly challenging with conventional, manual-oriented tools or simple automation. The sheer number of network elements, connected devices, diverse applications, and dynamic traffic patterns makes traditional manual network management insufficient. This becomes further complicated when a disturbance occurs in the network. AI and ML emerge as essential tools for harnessing the vast amounts of network metadata, enabling truly resilient networks. AI/ML techniques excel at predicting faults by analysing diagnostic data from network elements, allowing failures to be prevented before they happen. Additionally, these technologies enable proactive optimisation and rapid fault recovery, further strengthening network stability and performance. Key applications include predictive maintenance of base stations, which analyses telemetry data like temperature, power grid levels, and signal quality to schedule proactive repairs and avoid outages. AI-driven anomaly detection monitors traffic patterns to identify unusual spikes, potential attacks or impacts from natural disasters, enabling quick mitigation.

AI methods may help in realising self-organising networks, which will empower the network with self-diagnosis and possible repairing or healing strategies, allowing the system to detect faults and reroute traffic or reconfigure cells without human intervention, maintaining continuous service. Furthermore, dynamic resource allocation powered by ML optimises spectrum use, bandwidth, and power based on real-time demand and interference, enhancing service quality during peak times. AI facilitates network slicing optimisation by dynamically allocating virtual network resources tailored to specific service needs, ensuring critical services remain operational under stress.

With recent advancements in generative AI, the rise of AI agents might enable a new level of automation and intelligence in managing complex systems. These agents can autonomously analyse vast amounts of data, make decisions, and execute actions, transforming how networks are monitored, optimised, and healed to achieve greater resilience and efficiency. The agents can be optimised and trained through experience, such as by using deep reinforcement learning (DRL) and a digital twin of the network. This digital replica enables the simulation of various scenarios that challenge network functionality, allowing the agents to learn effective strategies for managing and improving network resilience in real-world conditions. We will discuss these in more detail in Chapter 5.

Simultaneously, AI methods enable predictive measures in other sectors intertwined with the mobile networks, such as the energy grids. For example, the energy grid stability may be affected due to fluctuating demand, decentralised generation, and the growing integration of renewables feeding energy into the grid. AI methods enable predictive load balancing by forecasting demand using historical data, weather forecasts, and real-time sensor inputs, allowing potential overloads to be identified before they occur. These systems can optimise energy dispatch by rerouting energy flows and engaging demand response to balance the grid proactively, which has a direct impact on the mobile network resiliency.

## 2.4 Interplay between mobile networks and other systems and sectors

Contemporary networked systems are becoming increasingly complex and interconnected. In addition to dealing with resilience engineering of networks operating in isolation, it is equally important to acknowledge that critical infrastructures are inherently closely coupled, and the proper functioning of one system depends heavily on the proper functioning of another (Buldyrev, et al., 2010). A typical example is the interdependency between the electric power grid and the communication network. Such systems of systems, although composed of highly resilient individual systems, may be considered less resilient as a whole (Buldyrev, et al., 2010). Interdependence of systems also raises the risk of cascading failures, e.g. in interconnected ICT and power systems (where ICT provides control services to the power system, while power system provides power supply to the ICT system), where even a single failure in one of these systems (e.g. of a power node) can trigger multiple failures in the other system, which, if propagate further, can result in a total collapse of both systems (Rak & Hutchison, 2020).

No longer isolated, 6G networks are deeply intertwined with external systems and cross-sectoral infrastructures. As we shall see next, building resilience into mobile networks requires a holistic and cross-sectoral approach.

### 2.4.1 Dependence on synchronisation systems

Mobile networks and other critical infrastructures, such as banking and energy systems, rely on GNSS for providing precise timing information. For mobile communications in particular, accurate timing is essential for coordinating base stations, especially in time division duplex (TDD) operations, and for enabling efficient handovers.

GNSS provide accurate timing that is readily available globally without any additional infrastructure other than a timing receiver. GNSS signals are relatively weak at the Earth's surface, thus susceptible to radio interference, including both unintentional disruptions and deliberate jamming. Jamming has increased in recent years, driven by geopolitical concerns, such as the use of GNSS jamming to deter drone operations, and individual motivations, including location anonymisation through self-jamming devices (which are cheap and readily available online).

The growing dependence on GNSS introduces systemic risks that extend beyond the telecommunications domain. In finance, even milliseconds of drift in synchronisation can disrupt trading systems. In energy grids, misaligned phase angles may lead to instability or outages. As threats like GPS spoofing and jamming become more sophisticated and widespread, ensuring timing resilience becomes critical for the operational continuity of multiple interconnected sectors.

Several possible mitigation techniques are under development or in early deployment, including

- interference-tolerant GNSS receivers utilising adaptive multiple antenna arrays (akin to multiple input multiple output (MIMO)) to suppress jamming signals.
- precision time protocol (PTP), also known as IEEE 1588, which uses a fixed network to provide timing and requires that network elements support this protocol.
- over-the-air (OTA) network time synchronisation using 6G signals, if properly adapted to contain the necessary timing information (Saarnisaari, et al., 2021).
- neighbour-reference timing strategies that utilise the nearest working GNSS timing point as a time reference, indicating that interference detection and integrity testing must be integral to the overall solution.

Finally, a hybrid, and preferably redundant, architecture that blends satellite, terrestrial, and OTA synchronisation mechanisms will be necessary. This multi-layered timing infrastructure should be standardised across sectors to prevent cascading failures in the event of GNSS degradation.

Moreover, robust GNSS interference detection and integrity validation between receivers must be embedded into system designs to enable timely mitigation and adaptive fallback.

Resilient timing should be embedded as a native feature of 6G networks, rather than treated as an auxiliary function, to ensure reliable operation across increasingly interdependent sectors.

### 2.4.2 6G, energy, and other sectors

6G networks depend on stable power delivery from the smart grid to operate energy-intensive components, including RAN, edge and cloud data centres, and core networks. The smart grid increasingly relies on 6G capabilities to enable real-time monitoring, predictive fault detection, and autonomous control, thereby mitigating cyber-physical threats. By leveraging URLLC and edge AI, 6G supports dynamic grid operations, enabling instantaneous load balancing and autonomous rerouting to mitigate disruptions. This mutual dependency underscores the vital role of resilience in ensuring business continuity and economic stability.

Real-world events illustrate how fragility in either system can trigger widespread societal and economic consequences:

- Chile's 2024 nationwide blackout halted copper mining operations, spiking global prices (Financial Times, 2024).
- In the United States, businesses lose $150 billion annually due to power interruptions, driving investments in microgrids and backup power (Department of Energy, 2017).
- Spain's April 2025 blackout caused widespread infrastructure and telecommunication disruptions, with an estimated €2 billion in productivity loss; the event was triggered by a sudden 15 GW loss in generation, possibly due to grid oscillations or interconnection failure (Kemene & Christianson, 2025).

Current resilience strategies often focus on isolated aspects (e.g., grid hardening or demand response) rather than holistic, data-driven approaches (Andersson, et al., 2022) (Hallegatte, et al., 2019). A unified framework must quantify resilience across the telecommunication and energy sectors, coordinate short- and long-term strategies for dynamic adaptation to evolving threats, and optimise their interdependencies. This ensures energy networks bolster 6G resilience, while 6G further improves grid resilience. Table 2.1 presents the operational measures for this unified approach.

*Table 2.1 Operational measures*

| Phase | Energy sector measures | 6G sector measures |
|---|---|---|
| Before event | - Grid hardening (e.g., underground cables) Predictive maintenance via AI microgrid deployment<br>- Renewable diversification | - Energy-efficient RAN design<br>- Backup power provisioning<br>- O-RAN adoption for flexibility<br>- Load forecasting |
| During event | - Autonomous islanding<br>- Dynamic load shedding<br>- Distributed energy resources (DER) activation | - Dynamic spectrum sharing<br>- Edge-based failover routing<br>- Battery-to-grid support |

| After event | - Rapid fault localisation<br>- Priority restoration for critical loads<br>- Post-event analytics,<br>- Root cause analysis and policy updates<br>- Infrastructure repair | - Network slicing for emergency services<br>- AI-assisted damage assessment<br>- Energy-aware traffic rerouting |
|---|---|---|

We foresee two strategies for optimising these interdependencies.

1. Optimising energy networks for 6G resilience:

- Renewable integration: while increasing renewables improves sustainability, their intermittency introduces operational uncertainties for RAN and edge/cloud infrastructure.
- Storage utilisation: deploying battery storage at cell sites can stabilise the power supply, ensuring uninterrupted 6G operations during grid fluctuations.
- Energy consumption analysis and algorithm development: Identify inefficiencies in 6G networks and opportunities for Open RAN (O-RAN) adoption to improve efficiency.

2. Optimising 6G networks for grid resilience:

- Flexible load management: shifting non-critical loads (e.g., caching, software updates) to off-peak periods reduces grid stress, using temporal and spatial flexibility.
- Energy storage for grid support: cellular network batteries can feed excess energy back to the grid during periods of peak demand, thereby enhancing reliability and generating revenue (Gholipoor, et al., 2025).
- AI-driven and renewable-aware resource allocation: ML models (e.g., reinforcement learning) can dynamically optimise computing, communication, and energy resources under renewable and supply-demand variability.

A future-proof resilience strategy must consider the telecommunications and energy sectors as interdependent systems. This requires strong engineering and coordinated regulation, as well as cross-sector simulation environments and shared resilience metrics.

## 2.5 The challenges ahead

Contemporary networked systems are becoming increasingly complex and interconnected. In addition to addressing the resilience engineering of 6G networks operating in isolation, it is equally important to recognise that critical infrastructures are inherently closely coupled, and the proper functioning of one system depends heavily on the proper functioning of another.

The dynamic structures envisioned in 6G systems, characterised by decentralisation, heterogeneity, modularity, and hierarchy, allow for emergent resilience. Hierarchy can evolve from the lowest level to create subsystems within a system, and it can be both vertically and horizontally loosely coupled. Vertical loose coupling provides time-scale separation, while horizontal loose coupling refers to weak coupling between modules. In the self-organising network hierarchy, modularity and local interactions will be utilised to complement vertical loose-coupling (Mämmelä, 2025). These principles will be crucial for the efficient, adaptive, and resilient operation of future 6G networks.

AI transcends the realm of a mere tool and assumes the role of an agent capable of profoundly influencing the balance of power. Its implementation could potentially favour centralised information networks, thereby increasing the likelihood of the emergence of authoritarian regimes. A democratic, distributed information network operates on a decentralised model. Most decisions are not made at a central hub but rather at more peripheral locations. Democracies possess robust self-correcting mechanisms that enable the identification and correction of errors in decisions made at the central level. To ensure the preservation of a decentralised distributed information network and mitigate the risk of excessive centralisation, a comprehensive suite of self-correcting mechanisms implemented through vertically and horizontally loosely coupled multi-agent systems is essential. Agentic AI systems are promising in such disconnected or resource-constrained settings, as they can act autonomously, collaborate with peers, and adapt to local conditions without relying on centralised control.

### 2.5.1 The efficiency, sustainability, and resilience trade-off

The vital role of communication networks in our society has underscored the importance of building future mobile networks upon three cornerstones: efficiency, sustainability, and resilience. Efficiency has evolved significantly from 1G to 5G, with each generation enhancing performance in various ways. Initially, the shift from analogue voice to digital multiplexing marked a crucial step. Following that, advancements such as wideband modulation and carrier aggregation improved utilisation and reduced the cost per bit. Most recently, the introduction of massive MIMO, densification, and software-defined RAN has increased spatial reuse and offered greater architectural flexibility. As we move towards 6G, this trend continues by extending efficiency from link-level bps/Hz to achieving end-to-end performance across radio, transport, core, and edge networks. However, unlike previous generations, 6G will not focus solely on efficiency. Instead, efficiency is co-designed with sustainability and resilience so that reducing resource consumption does not compromise graceful degradation, service continuity, or rapid recovery.

Sustainability has taken centre stage in support of a more sustainable world and in the design of environmentally conscious communication systems. A recent white paper highlights how sustainability has become an increasing focus within European Union (EU) smart networks and services (SNS) joint undertaking (JU) projects (Rezaki, et al., 2025). Generally, efficiency, sustainability, and resilience impose conflicting requirements. Balancing these objectives remains a key challenge in the evolution of communication networks. Resilience often relies on redundancy, backup systems, robust error correction, and other techniques that tend to increase energy consumption and hardware complexity. Efficiency aims to maximise performance per unit resource. Sustainability seeks to minimise resource usage, such as energy, materials, and infrastructure. Therefore, overly aggressive resource-saving measures may reduce a network's ability to respond swiftly to unexpected disruptions.

This tension is pronounced in the edge-cloud continuum architectures, where decentralisation enhances fault tolerance and resilience at the cost of an increased number of distributed components, which must be powered and maintained. Similarly, expanding coverage to underserved rural areas is essential for digital inclusion and resilience; however, this may involve higher per-user energy costs and the deployment of infrastructure in low-density regions. The dependencies on external systems further complicate the balance between these components. For instance, energy networking can help reduce reliance on disposable batteries at the edge, supporting sustainability; however, it also introduces new coordination and reliability challenges that must be managed as part of a resilient framework.

The flexibility and adaptability of AI, particularly distributed and multi-agent systems, are expected to play a key role in enabling attractive compromises between the often-competing demands of resilience and sustainability. Understanding these interactions between efficiency, sustainability, and resilience is crucial for identifying suitable engineering trade-offs.

# 3 Resilience Definition

Changes in the world around us, such as those driven by new applications and technologies, climate change, and disruptive events, bring new and often unforeseen challenges. Most notable among them are those related to malicious human activities, natural hazards, and technology-related system issues, such as hardware/software/human errors (Rak & Hutchison, 2020). Furthermore, the frequency, intensity, and scale of such disruptive events are notably rising.

In recent times, *system resilience*, *i.e.*, the ability to withstand, adapt to, and recover from such disruptive changes, has become a critical concern for system designers. The word **resilience** originates from the Latin word *resiliere*, meaning to rebound or spring back, and is defined as *the ability to recover to the original state after a disruptive incident*. In the social sciences, it is referred to as the psychological quality that enables some people to be knocked down by life's adversities and recover at least as strong as before. In business and organisation management, resilience is an organisation's ability to absorb stress, recover critical functionality, and thrive in altered circumstances.

The concept of resilience in the engineering domain is relatively new. In system engineering, it refers to the intrinsic ability of a system to adjust its functionality in the presence of disturbances and unpredictable changes. In the context of communication networks, resilience has been mostly considered from the perspective of security. For example, the International Telecommunications Union (ITU) has placed "security and resilience" at the centre of the network design process in its IMT-2030 (6G) framework (ITU, 2023). However, resilience is a much broader concept than just security. This chapter aims to provide a comprehensive definition of resilience, with a specific focus on wireless networks. Resilience is defined in terms of a generic system function, considering the complexity of systems and the interdependence of multiple systems. Benchmarks and metrics to measure its key performance indicators (KPIs) and key value indicators (KVIs) are also proposed.

## 3.1 Challenges leading to failures

Challenges in the operating conditions are an indispensable part of any system. Challenges are frequently responsible for **faults**, *i.e.*, flaws (imperfections) in the system, such as software bugs or hardware flaws, likely to occur at various stages of system engineering. A fault, if not properly dealt with, can lead to an **error**, defined as a deviation of the observed value/state from its specified (correct) value/state. Finally, errors are reasons for **failures** (Sterbenz, et al., 2010), *i.e.*, events occurring when the delivered service deviates from the correct service, see Figure 3.1illustrating the chain of discussed events.

> Figure 3.1
> Description: The chain of events leading to service failures

*Figure 3.1 The chain of events leading to service failures*

As illustrated in Figure 3.1, shortly after the onset of a failure event, the level of service may begin to decay from its nominal level, indicating that the service may no longer be provided. However, if its level is above the minimum acceptable level, service is still maintained. When the level of service delivery is below the minimum acceptable level, service is considered not to be maintained. The restoration of the affected service, e.g., by activating additional/backup resources such as backup servers/backup communication paths, may bring the service at least to the "maintained" state. A full recovery of all affected services often proves feasible only after the physical repair/replacement of failed network components.

Figure 3.2
Description: Phases of service recovery and network repair.

*Figure 3.2 Phases of service recovery and network repair based.*

The general concept of resilience is defined by the system characteristics at a point of failure, followed by a recovery. Therein, the detection of a disruption, remediation against the disturbance, and the process of recovery are pivotal for analysing system resilience. Suppose $s(t)$ is the desired service level at time $t$, which is the ratio between the service state and the minimum acceptable service level (as shown in Figure 3.2). For example, in a communication system, the service state could be the instantaneous data rate for a given link, while the minimum rate target could be the desired level. It is worth noting that $s(t) \geqslant 1$ implies the system meets the desired service level, while the opposite indicates a detection of a disruption.

Network resilience is built upon several challenge tolerance paradigms, such as:

- **survivability,** reflecting the system's ability to fulfil its mission promptly in the presence of faults (fault tolerance) and various threats, including attacks or natural disasters,
- **disruption tolerance,** referring to the ability of the network to tolerate disruptions in connectivity among its components, and
- **traffic tolerance,** associated with the ability to withstand the unpredictable (malicious/non-malicious) traffic (Sterbenz, et al., 2010).

The assessment of the ability of a networked system to deliver services (necessary to evaluate the degree of trust we can place in the system) is performed with respect to the system's **trustworthiness**, whose attributes are grouped into five areas, namely: **dependability,** referring to the level of reliance which can be placed on system services (involving attributes such as **availability,** denoting service readiness of its usage at a given time; **reliability,** referring to probability of service continuity; or **safety,** ensuring the non-occurrence of catastrophic consequences of the system's functioning on the environment) (Laprie, 1992); and **performability,** focusing on the system's performance in the context of the assumed Quality of Service (QoS) guarantees, and security, denoting the ability of a system to remain protected from

> Figure 3.3
> Description: Disciplines of network resilience from (Sterbenz, et al., 2010).

*Figure 3.3 Disciplines of network resilience from (Sterbenz, et al., 2010).*

unauthorised activities. The relation between disciplines and attributes of network resilience is explained in detail in (Sterbenz, et al., 2010) is illustrated in Figure 3.3.

## 3.2 Resilience definition: State of the art

The resilience and policy committees of the **National Academy of Sciences (NAS)** define resilience as a system's capacity "to prepare and plan for, absorb, recover from, or more successfully adapt to actual or potential adverse events" (Cutter et al., 2012) (Cutter et al., 2013). = This definition encompasses two essential and interdependent properties: system robustness (resistance) and system resilience. Traditionally, engineering has focused on designing systems with robustness, enabling them to withstand typical external and internal stresses. In contrast, resilience refers to the ability to spring back or rebound from unknown or unexpected disruptions. Similarly, in social-ecological thinking, resilience can be defined as the capacity to deal with change and continue to develop (Stockholm Resilience Centre, 2020)

Network resilience has primarily been considered in the context of security, referred to as cyber resilience (Lee, et al., 2022), which is a system's ability to prevent, withstand and recover from cybersecurity incidents. Historically, cybersecurity has emphasised building defences to prevent the loss of confidentiality, integrity, and availability in digital information and systems. However, in recent years, such a reactive approach has proven inadequate in the face of diverse cyberattacks. Cyber resilience has emerged as a complementary priority that seeks to ensure digital systems can maintain essential performance levels even when a cyberattack degrades capabilities. Thus, cyber resilience efforts frequently do not focus on whether an attack can happen and instead focus on how to react when they do occur (Lee, et al., 2022).

Although cyber resilience is an important part of network resilience, it does not fully encompass all aspects of system resilience. Resiliency is a much broader concept than just security and limiting its application in the context of cybersecurity does not fully detail the dynamic actions a network must take to guarantee protection, especially during major disruptions that may not be attacks, like natural disasters, cascading equipment failures, or failures stemming from interdependent systems such as the power network.

The development of wireless network generations generally prioritises higher data rates and lower latencies. However, emerging realities, such as the deployment of wireless networks in mission-critical use cases, a rapidly changing landscape, and increasing challenges, are driving a push toward resiliency, *i.e.*, networks that can better accommodate adversity and adapt to unforeseen challenges.

Communications networks are becoming an increasingly integral part of critical network infrastructure, serving mission-critical applications. For example, Finland's legacy public safety network will be replaced by the 4G/5G-based Virve 2.0 network provided by a telco. At the same time, the Future Railway Mobile Communication System (FRMCS), an international wireless communications standard for railway communication and applications, will be implemented based on 5G technology[2]. Therefore, these networks must continue to operate in the face of unknown disruptions.

Due to the indispensable nature of communication networks, resilience is needed as the core property of the network infrastructure (Hutchison, et al., 2023). Modern communication networks must be able to handle unknown and unforeseen events, both within the network and from external sources. This requires a holistic view of the resilience problem, leading to appropriate and easy-to-handle solutions.

Resilience, especially in the context of wireless networks, is not well defined, partly because telecom networks had not previously been able to predictably and reliably handle adversity. However, it is gaining traction in wireless research and standards, as highlighted by ITU in its IMT-2030 framework, where resilience is defined as the ability of a network or a system to continue operating correctly during and after a natural or man-made disturbance (ITU, 2023).

Resilience is often mistakenly considered synonymous with reliability or robustness. In essence, while reliability prevents failures and robustness resists expected challenges, resilience embraces the inevitability of failures and focuses on maintaining critical functionality despite them (Zissis, 2019). Early 5G systems often relied on layered combinations of robustness and reliability, reinforced by stringent service-level agreements tailored to specific use cases (Shafi, et al., 2017). However, such approaches are proving inadequate for the dynamic, interconnected, and mission-critical environments emerging in the 6G era. Therefore, resilience must become a primary design goal, not an afterthought. It uniquely addresses both short-term recovery and long-term adaptation, ensuring continuous service continuity, graceful degradation, and rapid restoration in the face of the unknown.

## 3.3 Resilience definition for wireless networks

In the context of wireless systems, (Khaloopour, et al., 2024) have formally defined a resilient system as one that is "is prepared to face challenges, withstand them, and prevent most from causing performance degradation. It can also absorb the impact of significant challenges, ensuring essential functionalities or a minimum service level. Moreover, it can recover (*i.e.*, short-term coping, bouncing back), adapt (long-term coping, bouncing forward), and evolve based on the experiences learned during this process."

---

[2] https://www.nokia.com/industries/railways/frmcs/5g-radio/

*Figure 3.4  Resilience concept based on generic functions (Heinimann, 2016).*

The discussion above motivates us to define resilience in terms of generic system functions (Heinimann, 2018) (Ungar, 2021), consisting of four biophysical functions: (1) resist within acceptable limits of degradation, (2) restabilise the crucial functions, (3) rebuild functionality up to a sufficient level, and (4) reconfigure flows of services and the enabling physical structures (Heinimann, 2018). These biological and physical capabilities are coupled with five cognitive functions: (5) perceive (detect and interpret) the state of the system and its environment, (6) understand its significance and meaning, (7) plan purposeful courses of actions and retrieve them from system memory or from a database, (8) release the most meaningful action, and (9) learn and adapt.

Figure 3.4 illustrates a resilience concept based on four biophysical and five cognitive functions, also indicating the temporal arrangement of the biophysical functions along the time axis (Heinimann, 2018). The y-axis shows the system performance. In particular, Figure 3.4 also indicates that during the event that starts at time $t\_pre$ and ends at $t\_post$, three biophysical functions – resist, re-stabilise, rebuild - are successively active. In contrast, attentiveness is a pre-event function, and reconfigurability is a post-event function. The 'during event' functions require acting on the short term and are assigned to be the inner loop of resilience. In contrast, the post-event functions – reconfigure, remember, and adapt – are often attributed the property of being outer-loop functions of resilience acting on a long-term scale (Sterbenz, et al., 2010).

Especially understanding the significance of the system state and its environment (5)-(7) during the event requires mindfulness sense-making. In general, mindfulness sense-making focuses on maintaining awareness of the present moment, encouraging individuals to pay close attention to what is currently happening. This process involves interpreting ongoing events, leveraging previous experiences, and existing knowledge to make sense of the situation and to initiate a useful course of action (Weick, et al., 1999). Because mapping ambiguous and uncertain events onto fixed mindsets and assumptions can lead to incorrect assessments and ineffective actions, mindfulness advocates moving away from inflexible, pre-determined responses in favour of more adaptive and context-sensitive decision-making.

From a resilience engineering perspective, the implementation of mindfulness sense-making in technical systems, in particular, and the cognitive functions (5)-(9) in general, is often termed **resilience-by-design.** At its heart, it is closely associated with the design of cognitive technologies, a core subset of AI. These are defined as technologies that enable machines to possess mental abilities, such as mimicking human behaviour, learning from experiences, and making decisions, ultimately infusing intelligence into non-intelligent machines. Research on mindfulness sense-making within the realm of AI is still in its very early stages and is, without doubt, the most challenging aspect in designing resilient systems. Especially when aiming for collaborative mindfulness sense-making, where individual AI instances are working together to understand current system states and jointly retrieve purposeful courses of action during events, this approach could lead to better decision-making and improve the resiliency of systems.

## 3.4 Wireless Network Resilience

Wireless networks are an integral part of the societal infrastructure and hence must be resilient to unforeseen natural and man-made disturbances. While attempts have been made to define wireless system resilience, its mathematical foundations are still underdeveloped. Unlike robustness and reliability, resilience is premised on the fact that disruptions will inevitably happen. This section aims to contrast resilience with related concepts of reliability and robustness, discuss the timeline of events before and after a disruption in the resilience framework, and introduce several metrics that measure network resilience.

### 3.4.1 Resilience vis-à-vis reliability and robustness

In the context of this paper, we acknowledge resilience, robustness, and reliability as closely related yet fundamentally distinct concepts, each addressing different facets of the wireless communication system under stress. Reliability and robustness are two key concepts that enhance the efficiency and performance of a system. Complex networks such as wireless networks and smart grids have emergence as a system characteristic, implying that the interactions and relationships between the components are dynamic, numerous and intricate, and cannot be predicted simply by understanding the components in isolation. Moreover, neither natural nor human-caused disruptive events could be prevented entirely. This requires improved effectiveness of a system via complex system design that can be measured as the combined effect of reliability, robustness, and resilience (Zissis, 2019).

**Reliability** is the probability that a system will perform satisfactorily and adjust to the demand and constraints of the system for a given period when it is used under specified operating conditions. This attribute evaluates the network performance in the event of a loss of one or two assets/components, typically accounted for in the design phase (Zissis, 2019). However, reliability assumes a relatively static environment and does not account for prolonged outages, cascading failures, or unforeseen disruptions. Reliability refers to the consistent delivery of service without failure over time, typically quantified using statistical guarantees, such as achieving 99.999% successful packet delivery. It emphasises failure prevention and smooth operation under nominal conditions. However, reliability assumes a relatively static environment and does not account for prolonged outages, cascading failures, or unforeseen disruptions.

**Robustness** builds upon reliability by emphasising the system's capacity to withstand anticipated disturbances and uncertainties. It is the measure or extent of a system's ability to perform the intended task under anticipated disturbances and uncertainties or faults in a fraction of its subsystems or elements not accounted for. This attribute considers the elimination of multiple assets and quantifies the network performance in the event of cascading failures (Koç, et al., 2014). Robust systems continue to operate regularly in the face of challenges or complete failure. It prepares the network to operate acceptably under worst-case conditions, such as known interference patterns or hardware degradations. Robustness is generally a static or design-time property, often resource-intensive, and assumes foreknowledge of the threats to be mitigated.

**Resilience**, in contrast, cuts across both reliability and robustness by focusing on the system's ability to anticipate, adapt to, recover from, and evolve through both known and unknown disruptions, including unexpected failures, cyberattacks, or natural disasters (Reifert, et al., 2023) (Reifert, et al., 2024). It is a dynamic, runtime property, drawing inspiration from biological systems through characteristics such as elasticity (returning to a functional state) and plasticity (reorganising structure to maintain functionality) (Karacora, et al., 2024). Resilience embraces uncertainty by enabling real-time adaptation, autonomous recovery, and long-term learning.

A recent framework adapted to wireless systems (Reifert, et al., 2023) quantifies resilience as a weighted combination of three metrics: the relative average performance degradation during the absorption and adaptation phases, and the time to recovery, normalised by a reference time. The current resilience frameworks in regulation and standardisation do not fully encompass the broader definition of resilience as "*the capacity to deal with change and continue to develop*," a concept rooted in social-ecological thinking (Stockholm Resilience Centre, 2020). This expansive, pro-innovation perspective is crucial to consider, as resilience and cybersecurity focus on threats, the cost of fault tolerance, and *business opportunities*. Resilience and security services can offer significant value and competitive advantages for companies within the telecommunications ecosystem.

### 3.4.2 Timeline of events in the resilience framework

Resilience in wireless networks is not a singular capability but rather a strategic and temporal continuum that spans actions taken before, during, and after disturbances and disruptive events. To effectively characterise resilience, we may distinguish between pre-event and post-event strategies, and to understand network behaviour along a spectrum of proactive, reactive, and adaptive responses (Karacora, et al., 2024) (Mahmood, et al., 2025).

As the names already suggest, pre-event strategies refer to actions, behaviours, and policies adopted by the network *before* any challenging condition occurs, while post-event strategies involve responses executed *after* such disruptions arise (Sterbenz, et al., 2010). Pre-event strategies focus on anticipation, preparation, and mitigation, and in wireless networks include, but are not limited to, resource over-provisioning, multi-connectivity, and redundancy (Karacora, et al., 2024). In contrast, post-event strategies address response, recovery, and adaptation, encompassing techniques such as damage localisation, system reconfiguration, service restoration, and learning from failures (Sterbenz, et al., 2010). For instance, after a base station failure, a resilient network might adapt its modulation and coding scheme, trigger autonomous re-clustering, or adjust transmission parameters to preserve service continuity (Reifert, et al., 2023). Both dimensions are critical. Over-reliance on pre-event design may render the system brittle in the face of unexpected stressors, while depending solely on post-event recovery can result in unacceptable service degradation or latency (Bennis, 2025).

### 3.4.3 Resilience mechanisms

Resilience mechanisms can be broadly divided into **proactive** (pre-planned) schemes designed to ensure resilience for certain failure scenarios assumed in advance (e.g. failures of single communication links by means of single backup paths installed before the occurrence of a failure) and **reactive** schemes where system resources (communication links, servers, etc.) necessary for recovery of affected services are set up for this purpose only after the failure occurrence (Rak & Hutchison, 2020). Proactive actions are taken before an event occurs, whereas reactive responses are triggered after an event has occurred. Proactive measures aim to reduce the likelihood and impact of potential failures through prediction, prevention, and strategic preparedness.

Although reactive schemes are far more resource-efficient than their proactive substitutes, they lead to slower recovery of affected services. Also, reactive schemes cannot ensure 100% service restorability (since network resources available for reactive recovery may not be efficient to restore all affected services). On the other hand, reactive measures focus on damage control and rapid recovery following a disruption. Pure proactive schemes are fast but resource-demanding. Also, proactive schemes cannot cover all possible failure scenarios (especially scenarios related to evolving failure regions).

In contrast to proactive and reactive actions, which are employed before or after an event occurs, adaptive strategies operate across both stages, enabling the network to evolve continuously. In such cases, **adaptive** methods are instrumental because they are partly based on proactive mechanisms configured in advance and then, reactively adjusted to the actual characteristics of the failure scenario (Mahmood, et al., 2025). Adaptive systems adjust dynamically based on current conditions and past experiences, with a strong emphasis on learning and generalisation to better handle future disruptions. Concrete examples include resilient-by-design architectures as proactive strategies (Khaloopour, et al., 2024), event-triggered resilience mechanisms as reactive responses (Karacora, et al., 2024), and world-model updates and reasoning as adaptive capabilities (Bennis, 2025).

Resilience mechanisms and enablers, in terms of individual technology components and network architecture, are detailed in Chapters 4 and 5, respectively.

## 3.4.4 Resilience metrics

A central challenge in applying this framework lies in selecting an appropriate performance metric. Common choices in wireless systems, such as spectral efficiency, power consumption, or reliability, may not be well-suited for evaluating resilience. This issue remains open and calls for further research. Additionally, the framework is challenging to apply analytically, making it most suitable for large-scale numerical evaluation. However, given the extremely low probability of typical resilience events, standard Monte Carlo simulations may prove inadequate, underscoring the need for resilience-specific simulation techniques.

Measuring the resilience of a (wireless) communication system remains a largely unsolved problem. A recent framework adapted to wireless systems in (Reifert, et al., 2023) quantifies resilience as a weighted combination of three metrics, inspired by the recovery curve shown in Figure 3.5. These are the relative average performance degradations during the absorption and adaptation phases, as well as the time to recovery, normalised by a reference time.

There could be several steps towards recovery and return to normal operating state once a disruption occurs. It may be in discrete steps or as different continuous processes, as illustrated in Figure 3.5 (left and middle, respectively). There are three distinct recovery phases in a discrete system state case, namely absorption, adoption, and recovery. Then, the overall system resilience is defined as the weighted sum of functions of the time spent in each phase, where the weights for each phase add up to one.

Resilience in the case of a continuous system state is quantified with the recovery curve (Sharma, et al., 2018). It can be quantified using the dynamics of the recovery process as the normalised area under the system state during the recovery process, as shown by the shaded regions in Figure 3.5 (middle).

However, this definition does not distinguish between different recovery processes. For example, the two different continuous recovery processes in Figure 3.5 (middle) both have the same resilience metric, despite being distinct. As a remedy, the cumulative resilience function (CRF) is defined in (Sharma, et al., 2018) as a term analogous to the cumulative density function in probability theory. The CRF, as illustrated in Figure 3.5 (right), describes the system state during the recovery process, while its time derivative yields the instantaneous rate of the recovery progress, enabling the calculation of the recovery process over any given time interval. Hence, these two functions enable a more detailed analysis, thereby allowing for a comparison between the resilience of two different recovery processes.

Figure 3.5
Description: Illustrative example of a system's recovery process after failure through remediation with multiple recovery steps for discrete recovery steps (left), and multiple recovery mechanisms for continuous recovery steps (middle), along with their corresponding CRFs (right).

*Figure 3.5 Illustrative example of a system's recovery process after failure through remediation with multiple recovery steps for discrete recovery steps (left), and multiple recovery mechanisms for continuous recovery steps (middle), along with their corresponding CRFs (right).*

One major shortcoming of this approach is its neglect of failover strategies. In large wireless networks, failures in individual components can often be absorbed by rerouting functionality to other nodes, frequently with minimal performance loss and little time to recovery. Building on this observation, (Ahmadian, et al., 2020) proposed the Component Resilience Index (CRI), which quantifies the resilience of a component in terms of its criticality to overall system functionality, explicitly accounting for failover potential. This is a promising direction for evaluating resilience in wireless networks, but it shares the same limitations regarding metric selection and analytical tractability.

An alternative path is to acknowledge that resilience is inherently multi-faceted, making the pursuit of a single scalar metric both difficult and perhaps overly reductive. Instead, resilience could be assessed through a set of key indicators (Heinimann, 2016), such as reliability, survivability, time to recovery, diversity, and detection capabilities. This approach builds on established work in these areas while embracing the complexity of resilience. It also offers a practical benefit: some aspects of resilience can now be quantified, even as others remain elusive. Chief among the latter are behavioural aspects such as learning, self-evolution, and autonomous behaviour. These cognitive functions are central to resilience, yet no established methodology exists to quantify them in technical systems.

The R-value concept is an alternative generic resilience metric (Mieghem, et al., 2010). It is a linear combination of several graph metrics that quantify resilience in networks, such as average shortest path length, diameter, and assortativity, as well as more advanced metrics, including algebraic connectivity and spectral radius. Recently, the R-value concept has been extended to solve two open issues, namely, how to dimension several metrics to allow their summation, and how to weight each of the metrics (Manzano, et al., 2014). The (enhanced) R-value will be used to define several resilience classes. A resilience class specifies, for a particular service, a subinterval of [0, 1] since R ∈ [0, 1]. For example, class C1 contains all graphs whose R-values lie between [0, r1], class C2 contains all graphs in [r1, r2], and so on. The rationale behind resilience classes is that a small number of classes is more manageable than a continuous range of R values, and they ease interpretation by mapping the R values to a few ranges, such as red, orange, and green, with their usual meanings.

# 4 Architectures for resilience

Faulty links, compute nodes, timing lines, or power feeds can cascade outward in today's tightly interwoven mobile systems, turning a local glitch into a network-wide outage. The traditional response, namely over-provisioning monolithic subsystems and bolting on protection features later, is inefficient, hard to scale, and difficult to manage, leaving MNOs with rigid recovery options and rising costs. These shortcomings tilt the scale toward a resilience-by-design philosophy for 6G with an integrated network architecture, as illustrated in Figure 4.1, with users, access, compute and control, and orchestration levels.

> Figure 4.1
> Description: An overview of the integrated 6G architecture. It illustrates the different layers of an E2E 6G network, incorporating additional functional blocks.

*Figure 4.1 An overview of the integrated 6G architecture. It illustrates the different layers of an E2E 6G network, incorporating additional functional blocks such as time synchronisation and energy resources, which are relevant across all layers. Resilience enablers, including redundancy, reconfiguration, telemetry, and autonomy, are also highlighted, demonstrating their roles throughout the architecture to support robust and adaptive network operation.*

Four architectural levers set the pace of adversity detection, absorption, and rebound:

- *Where does redundancy live?*
- *How are failovers orchestrated?*
- *What is telemetry observed and how?*
- *How much autonomy is pushed to the edge?*

Architecture design choices, such as network functions placement, transport technologies, edge-cloud intelligence splits, and exposure of energy or timing sources, exercise such levers, framing our discussions in this chapter. Specifically, we first outline the principles that integrate radio, transport, compute, energy, and timing into an E2E resilience fabric, with the RAN serving as the anchor point. Then, we discuss how deep programmability and AI-native control reinforce these principles, drilling down to edge-hosted, private, and device-level scenarios, expanding to multi-domain and hybrid access, and concluding with distilled insights for building resilient 6G networks.

## 4.1 Resilient architecture foundations

6G networks must push further for lightweight central coordination (policy, intent, guardrails, cross-domain optimisation) combined with fast, local control loops at the edge. This is because purely centralised designs concentrate risk and add control-loop latency, while fully distributed ones shorten feedback but trade some efficiency and controllability. The EU high-voltage grid illustrates this balance, with market scheduling and cross-border set-points decided centrally on min-to-h timescales (i.e., centre setting intent), as well as sub-second protection and frequency control operated locally to isolate faults and keep islands stable (i.e., local loops acting autonomously, confining disturbances and restoring service efficiently).

The unifying principle is loose coupling (Mammela et al., 2023), such that subsystems' interaction limits fault propagation and permits graceful degradation. This requires designing for weak interdependencies, separation of timescales, and well-defined fallbacks, so that when conditions deteriorate, the system naturally reverts to simpler, decoupled modes without relying on brittle central coordination. This does not come for free, with extra costs arising from softwarized components and (buffering) redundancy and diversity reserves.

### 4.1.1 End-to-end network architecture

A network's resilience is bounded by its parts (see Chapter 3). A service can never be sturdier than the weakest indispensable block/component along its delivery chain. Whenever a function or link is replicated, e.g., dual CU instances in separate availability zones or two parallel LEO gateways, the aggregate resilience becomes the probabilistic combination of the replicas rather than the minimum of their individual scores. We therefore differentiate between critical-path elements, which lack redundancy and must remain operational, and replicated elements, whose diversity absorbs faults and ensures continued operation. In both cases, the multi-level principle still applies: a layer's robustness forms the platform on which the next layer builds, but the overall metric is now path- and criticality-dependent, rising with each added diversity layer yet collapsing if any irreplaceable component fails (Bennis, 2025) (Paul Smith et al., 2011).

6G E2E architectural resilience is the capacity to maintain "acceptable" service levels across the entire 6G service chain. Each layer or service chain component has its critical role. Indeed, centralised clouds provide scale, regional cores host latency-tolerant functions, and near-edge sites run time-sensitive CU/DU procedures. This continuum is stitched together by xhaul links that the RAN can observe and reconfigure, turning transport from a hidden dependency into an active resilience lever. This also requires exposing interfaces to other critical infrastructures. In practice, the RAN surface increasingly provides southbound APIs toward smart-grid energy controllers (e.g., IEEE 2030.5) and time-as-a-service providers (e.g., NTP over TLS). Such hooks proved vital during the 2025 Iberian blackout,[3] when Vodafone shed non-critical carriers and switched to hold-over oscillators, sustaining emergency calls for seven hours on generator power.

We can distinguish two key perspectives: i) resilient network infrastructure (i.e., radio, transport, compute, timing, and energy) itself, and ii) resilient services running on top (i.e., slices and applications with mixed criticality). In both cases, resilience must encompass not only engineered SLAs for anticipated faults but also emergent responses to unforeseen conditions.

---

[3] https://www.capacitymedia.com/article/spains-telecom-networks-run-on-backup-power-as-iberian-peninsula-goes-dark

Figure 4.2
Descriptions: Pillars of 6G Network Architecture Resilience

*Figure 4.2 Pillars of 6G Network Architecture Resilience.*

E2E design pillars are (see Figure 4.2):

- Modularity & disaggregation: fine-grained CU/DU/RU splits and micro-services contain failures and enable targeted restart (Hoffmann et al., 2023) (Mammela et al., 2023) (Reifert et al., 2024).
- Redundancy & diversity: multi-connectivity, multi-vendor open interfaces, and path diversity absorb localised damage (Paul Smith et al., 2011)(Reifert et al., 2024).
- Programmability: SDN data-plane agility plus intent-driven orchestration accelerates recovery actions (Hoffmann et al., 2023)(Khaloopour et al., 2024)(Mammela et al., 2023).
- Observability: pervasive telemetry and digital-twin replicas of the RAN create a live "nervous system" for anomaly detection and what-if simulation (Masaracchia et al., 2023)(Paul Smith et al., 2011)(Reifert et al., 2024).
- AI-native design: near-real time (RT) and non-RT RICs learn normal baselines, predict failures, and coordinate cross-layer mitigation while loose-coupling principles keep control loops stable (Khaloopour, et al., 2024)(Masaracchia et al., 2023)(Reifert et al., 2024).
- Sustainability: weighing economic, societal, and environmental aspects (Hoffmann, et al., 2023)(Lopez, et al., 2023)(Paul Smith, et al., 2011).

These act as "multipliers" for the resilience curve to rebound faster and more sustainably than any isolated block could on its own. Current deployments somewhat combine these principles, but challenges such as synchronising state across thousands of dispersed CUs, securing zero-trust handoffs between orbital and terrestrial segments, and sustaining SLAs when backhaul or energy budgets fluctuate persist (Mammela et al., 2023)(Masaracchia, et al., 2023).

### 4.1.2 RAN considerations

The RAN is the frontline of service delivery and the most exposed and capital-intensive part of the network (Masaracchia, et al., 2023). Hence, it is a critical domain for architectural resilience. Toward 6G, loose coupling within RAN control loops is central. Orthogonality in time, frequency, or space remains the default mechanism for isolating subsystems, while separation of time scales ensures stable coordination across layers.[4] However, the aim is not strict orthogonality or complete separation, but to ensure coupled subsystems degrade gracefully into decoupled fallback modes without active control. A practical example is the use of soft states, which expire automatically if not refreshed, providing an alternative to fully stateless designs. Meanwhile, network slicing and resource isolation further strengthen fault containment; however, architectural safeguards are still required to prevent slice-level disruptions from compromising the baseline functions of the RAN.

A second axis of resilience concerns the placement of functions. Centralised RAN architectures benefit from coordination efficiency and cost-sharing, but when connectivity to the core is interrupted, their dependence on centralised control becomes a liability. In contrast, edge-hosted RAN functions allow local survivability during backhaul outages, supporting scenarios such as disaster recovery. Future architectures will require the flexibility to dynamically shift CU/DU functions between the core and edge, striking a balance between efficiency and autonomy. Related to this is the RAN's ability to operate in limited-capacity modes. Instead of a complete outage, degraded operation with reduced bandwidth, lower QoS, or best-effort service provides critical continuity. Applications, in turn, must be designed to adapt gracefully to such degraded modes, preserving essential communications while suspending non-critical features. Different services must degrade coherently rather than fragmenting into incompatible behaviours, which is essential when slices span diverse RATs or when fallback paths include non-3GPP domains.

Infrastructure mobility/dynamicity is another new pillar. Unlike earlier generations that assumed static deployments, 6G must accommodate nomadic networks, vehicular BSs, and UAV-mounted RUs. These mobile infrastructures can be rapidly deployed to restore service in disaster areas or to augment capacity in temporary hotspots. While they create new challenges in authentication and orchestration, they also redefine the resilience toolkit by making coverage and capacity mobile assets.

## 4.2 Programmable and AI-native networks

Programmability and AI are central to 6G resilience as they provide the means for dynamic and intelligent auto-reconfiguration to avoid and/or recover from adverse events. Unlike 5G, where these capabilities were added, 6G is expected to incorporate them during the design phase. This means embedding programmable functions and learning loops directly into the E2E architecture (Hoffmann, et al., 2023). The goal is to create networks that are not only adaptive but also self-evolving and self-organising.

---

[4] Time-scale separation between control layers (with approximately RT <10ms, near-RT: 10ms–1s, non-RT >1s) provides an orthogonality degree, though boundaries overlap in practice and MNOs engineer explicit guard-bands and coordination. ETSI's GANA framework adds a useful perspective by mapping distributed 'decision elements' to RIC xApps/rApps (Mammela et al., 2023), helping formalise stability across time scales.

## 4.2.1 Programmable networks

5G programmability was introduced via SDN, NFV, and the first wave of disaggregated RAN deployments. O-RAN has extended this vision with near-RT and non-RT RICs, exposing interfaces where control logic can be added, upgraded, or reconfigured. This allows modularity and partial automation, but programmability remains fragmented. Functions are often confined to isolated domains, and vendor-specific APIs still dominate; orchestration across RAN, transport, and core remains immature. For instance, programmable RAN functions have so far mainly targeted narrow optimisations, such as traffic steering, anomaly detection, predictive maintenance, or energy saving, rather than providing critical mission support, robust recovery, or system-wide fault containment. A significant challenge is that while virtualisation and cloudification enhance flexibility, they also enlarge the attack surface and risk propagating faults across domains if policies or updates are misconfigured. Compromised VNFs, faulty containers, or unverified updates may trigger cross-domain failures.

Advancing toward 6G requires turning programmability from a set of domain-specific knobs into an E2E capability that is both standardised and resilience-aware. This means unified intent-based orchestration across RAN, transport, and core, with a common abstraction layer for APIs that eliminates vendor lock-in. Programmability also opens new modes of resilience through multi-purpose reuse of infrastructure. Indeed, since dedicated sensing modules are unlikely to be widely deployed in future RANs due to cost and limited utility in peacetime, the integrated sensing and communication paradigm is receiving considerable attention. For example, O-RAN DU with carrier sweeping may achieve intrusion detection with very low capacity overhead, enabling programmable multi-modal awareness.

Wireless transport must also evolve into a first-class programmable component, capable of energy-aware rerouting, latency-bounded guarantees, and dynamic path selection across fibre, microwave, NTN, and potentially THz-based mesh backhaul. In fact, current WAB solutions can fill gaps when fibre is absent (subject to latency, throughput, and energy efficiency constraints), but this has still been done without designing with resilience as a primary goal/requirement (Khaloopour et al., 2024). The goal is to transform monolithic network functions into modular, software-defined components that can be orchestrated across a distributed infrastructure.

To avoid programmability itself becoming a liability, mechanisms such as sandboxing for new functions, automatic rollback on instability, and zero-trust enforcement at the interface level must be embedded into the architecture. Without such safeguards, programmability risks amplify faults rather than absorb them. Notably, technology providers promote programmability as a differentiator, promising agility and multi-vendor diversity, but many MNOs remain sceptical (see Chapter 6). For them, programmability often increases operational complexity and staff training requirements without yet showing clear OPEX savings; hence, they still consider it more like a pilot feature than a mission-critical tool. Accelerating programmability adoption requires broadly demonstrating predictability/forecasting features and robust fallback mechanisms, and lowering costs.

### 4.2.2 AI-native networks

Programmability provides the means for dynamic reconfiguration, while AI must decide "how", "what", "when", and even "why" reconfiguration should occur. The 2022 Rogers Communications outage, triggered by a router misconfiguration and affecting over 25% of Canadians, including emergency calls and financial transactions, clearly motivates the need for dynamic reconfigurability. However, contemporary automation/AI alone is not exempt from issues. The 2023 Optus outage incident evinced this when automated network protection mechanisms, triggered by a surge of routing updates, instructed core routers to withdraw routes and disconnect, cascading into a nationwide collapse that left over 10 million Australians without connectivity.[5] These cases demonstrate that purely human or purely automated approaches are insufficient, underscoring the need for hybrid human–AI governance.

Towards 6G, it is expected that near-RT RICs will run AI-based schedulers, non-RT RICs will manage longer-horizon policy optimisation, and distributed edge nodes will contribute to federated and privacy-preserving learning, as well as operating semantic communication protocols that optimise control information exchange. This allows the network to anticipate faults, proactively adapt resources, and coordinate recovery across the RAN, transport, and core, thereby realising self-healing and self-optimising infrastructures. For instance, an AI-native framework jointly orchestrates slices across domains, showing up to 10% higher slice success rates and 20% faster recovery compared to reactive benchmarks. However, MNOs have so far remained reluctant to entrust AI with real-time, safety-critical decisions due to limited explainability, the lack of large-scale datasets for training, and the absence of regulatory and operational frameworks to ensure SLA compliance. These issues remain actively under investigation for mitigation.

For AI to truly contribute to resilience, it must be resilient itself. Models can drift, be poisoned, or diverge in training, and AI agents may deliver unsafe recommendations if unchecked (Khaloopour, et al., 2024). Containment strategies, therefore, need to be architectural, including fallback policies that revert systems to deterministic defaults when AI fails, continuous validation and monitoring, and human-in-the-loop oversight for exceptional situations (see Chapter 5). AI-native networks also require broad situational awareness to operate safely, which involves both extensive introspection (via telemetry and digital twins) and environmental awareness, e.g., via RF (and multi-modal in general) sensing, as mentioned earlier. Finally, we emphasise that resilience requires designing AI-native networks with built-in containment and recovery mechanisms.

---

[5] https://www.reuters.com/technology/australias-optus-hit-by-national-network-outage-2023-11-07/

## 4.3 Edge-native and localised networks for resilience

Edge-native systems are crucial for delivering certain service levels during adverse conditions, isolating disruptions locally, and accelerating recovery through distributed adaptation. A transition from cloud-centric to edge-empowered infrastructures within the edge–cloud continuum is desired. The localised capabilities may not replace centralised intelligence but complement it, creating hybrid loops where global intent and policies are refined by local sensing and decision-making.

### 4.3.1 Edge hosting of RAN functions

A key enabler of edge-native resilience is the migration of selected RAN functions from centralised cores toward MEC sites and distributed edge nodes. Note that centralisation of control and user functions has so far offered efficiency through pooling and simplified management (Masaracchia et al., 2023) (Reifert et al., 2024). However, it has also increased dependencies. For instance, when backhaul to the core is lost, BSs that rely on central authentication or mobility anchors may struggle to sustain service, unless limited fallback modes are supported locally. This tension motivates the shift toward edge-hosted RAN functions in 6G, where autonomy at the edge can complement centralised pools and reduce the risk of service disruption.

MEC integration allows RAN functions to be co-located with application logic and data-plane services, reducing recovery times and enabling direct cross-layer optimisation. For instance, edge nodes bring together near-RT RIC functions, storage, and distributed intelligence in ways that enable a range of localised resilience mechanisms. These include:

- caching and replication, allowing critical content to remain accessible during transport failures;
- local decision-making, i.e., by edge RICs or DU servers, without round-trips to the core;
- replication of control/user plane functions, such as dual CU instances across edge zones, which absorb localised failures.

Edge hosting also deepens the role of distributed intelligence, while programmability and AI loops embedded at the edge enable proactive reconfiguration and distributed learning across local clusters, allowing for the anticipation of stressors and real-time resource allocation.

Critical challenges for resilient, edge-native architectures include i) ensuring state synchronisation across dispersed CU/DU replicas (Hoffmann, et al., 2023) (Reifert, et al., 2024), ii) enforcing zero-trust security for autonomous edge nodes, and iii) avoiding unsustainable

OPEX growth from distributed infrastructure. The latter two aspects are revisited in Chapters 5 and 6, respectively.

### 4.3.2 Transport and synchronisation dependencies at the edge

While distributing functions closer to the access layer increases autonomy, it does not eliminate the vulnerabilities of transport and synchronisation. Transport resilience relies on programmable xhaul with redundancy. Meanwhile, complementary timing sources such as PTP hierarchies, multi-source time feeds, and mesh synchronisation are essential to address the single point of failure of GNSS for global synchronisation (see Chapter 2). At the edge, these mechanisms allow terrestrial BSs, LEO relays, and HAPS nodes to form quasi-independent clusters, which can run autonomously on holdover oscillators, with stability ranging from seconds (low-cost quartz) to hours (rubidium or chip-scale atomic clocks) before exceeding the tens-of-$\mu$s accuracy thresholds required for air-interface synchronisation. Upon recovery of backhaul or external timing, resynchronisation can be achieved, provided that jitter accumulation during holdover is minimised.

The nature of disruptions dictates the required tactics. Indeed, extended outages, such as fibre repairs or GNSS jamming lasting hours, may require reallocation of functions across clusters and controlled re-parenting of timing hierarchies. In contrast, short and irregular disturbances, such as packet delay variation in PTP flows, require fast re-sync strategies, often at sub-s timescales, using local reference nodes or boundary clocks. Mesh synchronisation across edge nodes offers further resilience but must be engineered to limit error propagation, for instance by bounding per-hop jitter to tens-of-ns when supporting URLLC slices. Meanwhile, beyond redundancy, xhaul and timing systems can fall back to limited-capacity modes. For example, fronthaul links may downgrade to lower modulation/coding and insert larger guard periods to tolerate sync uncertainty, while applications are notified of reduced service. Such adaptive degradation ensures continuity for critical traffic, even if throughput is curtailed.

All in all, survivability at the edge depends on a combination of domain isolation, transport redundancy, and adaptive degradation.

### 4.3.3 Survivability domains and resilience islands

Private and localised RAN deployments are nowadays a natural complement to public mobile infrastructures, particularly in vertical domains such as manufacturing, factories, healthcare, energy, defence, and public safety (Guo, et al., 2022). They are bound to a site, operated by an enterprise, and architected to sustain local operations independently of public infrastructure, making them designed for "survivability domains".

Private 5G RANs already support on-premise compute and storage, typically through co-located MEC servers and local cores, enabling intra-site services to continue during external outages. What remains underdeveloped is the ability to sustain truly deterministic and resilience-oriented services at scale. For example, 6G will need to enhance TSN integration for industrial control, expand support for critical healthcare applications such as telemedicine, and standardise survivability modes so that such services can continue with limited but sufficient scope even under long-lasting disruptions.

Meanwhile, "resilience islands" denote opportunistic, *ad-hoc* RAN clusters emerging when public networks are disrupted. Here, the goal is not business continuity for a single enterprise, but rather to maintain acceptable communication services for users in the affected area. Tactical "bubbles" used by public safety actors during disaster response illustrate this principle: mobile systems combining RAN and core in a single portable package provide local connectivity for responders, and opportunistically re-attach to national cores when transport becomes available again. While such self-contained bubbles are feasible for authorities, extending them to *ad-hoc* public resilience islands raises unresolved questions about user authentication, security, and governance. Current architectures do not support transparent, trustworthy user admission without core connectivity; hence, MNOs are reluctant, as there is no way to enforce billing or identify paying subscribers.

The above concepts are closely related to the notion of "*sub-networks*" (Hoffmann, et al., 2023), where clusters of RAN nodes and edge compute can continue operating autonomously when disconnected from the MNO core. In this framing, private/campus networks represent planned sub-networks designed for enterprise survivability. In contrast, resilience islands represent *ad hoc or emergent subnetworks* that form under disruption in public domains. See a side-by-side comparison of these two technologies/concepts in Table 4.1.

The resilience value of the subnetworks lies in autonomous operation and detachment from centralised control. Indeed, local orchestration enables caching, replication, and scheduling to continue without involving the cloud or core, while local edge compute sustains critical functions. However, significant challenges remain, especially in terms of i) security and trust, as zero-trust principles imply that even isolated clusters must authenticate and authorise users reliably, but existing mechanisms often depend on the central core;[6] ii) algorithmic and engineering limitations with alternative xhauls like LEO and multi-hop wireless, since current PHY/MAC designs cannot tolerate much higher latency and jitter than what they are primarily designed for; and iii) governance and interoperability, raising questions of who operates, secures, and regulates these domains/islands and how (see Chapter 6).

---

[6] Significant overhead will arise from continuous certificate validation and additional signaling needed to manage credentials locally.

*Table 4.1: Survivability Domains vs Resilience Islands*

| | Survivability Domains (Private RANs) | Resilience Islands |
|---|---|---|
| **Scope** | enterprise/campus sites (e.g., factories, hospitals, ports, utilities, military bases) | ad-hoc network clusters |
| **Operator** | enterprise, vertical, or dedicated service provider (closed to outsiders) | opportunistic. It could be MNOs, public authorities, or dynamically by available RAN equipment. |
| **Design intent** | business continuity and critical service assurance. | emergent fallback for "acceptable" connectivity under adverse events |
| **Resilience role** | maintains critical intra-site communications and deterministic services during WAN/core outages | keeps users and responders in the affected area connected locally, supports emergency coordination until the wider network returns |
| **Key challenges** | smooth interoperability with public networks, lifecycle management, security integration, CAPEX/OPEX for enterprises | lack of standardised authentication/ billing without core, unclear governance/regulation, technical immaturity of *ad-hoc* xhaul |
| **Key enablers** | MEC, TSN support, local orchestration, caching/replication, zero-trust access, integration with enterprise information/computer department | autonomous control loops, latency-tolerant PHY/MAC for alternative xhaul, multi-source timing, local authentication |
| **Maturity** | medium: commercial deployments underway in 5G, early vertical use cases and local survivability proven, integration still evolving) | low: mostly conceptual, limited to tactical bubbles (e.g., military, public safety), requires significant research and development, and new governance frameworks |

## 4.3.4 Far-edge resilience

Massive-IoT swarms and mission-critical sensors push 6G to manage billions of resource-constrained devices that must operate under harsh conditions, e.g., limited resources, poor channels, timing/localisation gaps, and vulnerability to physical tampering (Lopez, et al., 2023) (Nurul H. Mahmood, et al., 2020). While redundancy in the form of multi-connectivity, large batteries, and/or standby backhaul links can improve resilience, such measures are not always feasible, cost-effective, or sustainable, motivating architectural tactics for alternative or additional resilience support (Lopez, et al., 2023) (Paul Smith, et al., 2011). Indeed, true far-edge resilience requires architectural support. That is, structuring the system so that recovery and adaptation mechanisms can be invoked automatically, safely, and at scale (Mammela, et al., 2023)(Paul Smith, et al., 2011). For this, we identify the following three architecture-level building blocks.

- **Edge-resilience control layer**: A tier of gateways and aggregation hubs may host a local control plane that buffers traffic, runs tiny-AI models, and stores scripted fail-over playbooks. This layer must quarantine adverse events, throttle non-critical traffic, and execute recovery scripts promptly, ensuring that safety or telemetry data continues to flow. For instance, if a sensor or gateway goes silent, the layer can flag the fault, recompute the local cluster map, and issue re-clustering or task-migration orders so nearby nodes with overlapping coverage take over the missing measurements or forwarding role. Devices may periodically broadcast health hints, such as voltage/energy levels, queue depth, and sensor sanity checks, to feed predictive fault models (López, et al., 2025). Even in adverse conditions, the network should be able to interact with devices to coordinate actions, e.g., by suspending transmission attempts, such as RACH packets, for energy conservation, and/or devices must include proper primitives to handle unforeseen conditions. These are crucial to enhance their action set possibilities in the near future and their long-term resilience.

- **Adaptive connectivity fabric**: Beneath the control layer, devices and gateways may exploit a unified "any-RAT" fabric capable of routing packets over multiple RATs or opportunistic mesh relays. Indeed, if xhaul or a sibling gateway dies, policy engines may select alternative routes on the fly, balancing energy, latency, and channel quality. For instance, an edge RIC noticing battery drain can reroute traffic through a better-powered neighbour or spin up an LPWAN micro-slice, sustaining essential data until full-service returns.

- **Local trust & resource guardian**: An edge-resident guardian service may authenticate nodes and replacements, e.g., using RF fingerprints or cached credentials when cloud PKI is unreachable, flag spoofing attempts or jamming, and monitor network resources (including energy) to steer workloads toward resource-rich paths. The guardian can isolate suspicious nodes, recruit temporary relays (e.g., UAVs), and steer workloads toward energy-rich paths, enabling the cluster to absorb adversities gracefully.

An energy networking architecture can help support the above functions, and may constitute an important subsystem for far-edge resilience. Indeed, active energy distribution, where nodes with surplus energy share with deficit peers, or where access points and beacons provide wireless energy transfer (Lopez, et al., 2023), can help smooth imbalances and mitigate fragile/weak spots, and/or help recover from energy outages quickly.

The resilience of edge computing systems is inherently tied to the availability of energy. Devices at the edge, particularly IoT devices, rely on batteries; therefore, it is essential to ensure that battery operation does not compromise the resilient operation of these devices. This is especially important for edge computing architectures discussed in Section 2.3, where decentralised processing demands robust, uninterrupted power sources. Operation interruptions due to battery depletion and battery replacements during maintenance are examples of situations where the system's resilience is compromised. Nevertheless, battery-less devices, powered via energy harvesting, would be preferred, as they support sustainability, a key characteristic of the future 6G (Perera, et al., 2025). Consequently, the operation and performance of these energy-autonomous devices depend on the energy harvested from the ambient environment. Typically, energy is harvested from light and RF sources. This energy harvesting process is dependent on, among other factors, the position of the nodes. Nodes in good harvesting positions, e.g., located right below a lighting source, will be able to harvest enough energy to operate, and eventually, will have excess energy (Katz & Perera, 2023). Energy networking has been proposed to ensure that all zero-energy nodes in a network receive sufficient energy to operate, regardless of their location. It nodes enable nodes with excess energy to share energy with nearby nodes, using energy links that are implemented in the optical RF or inductive domains. Ultimately, energy networking envisions remote charging of batteries to eliminate the need for maintenance due to battery depletion. While energy networking enhances resilience at the device and edge, resilience at a larger scale requires coordinated interaction between telecommunications and energy systems. Next, we examine how 6G and energy systems can mutually strengthen each other's resilience.

## 4.4 Multi-access integration

6G resilience requires seamless integration of multiple access technologies and domains to realise architectural redundancy and diversity by default or on demand. Indeed, TNs, non-NTNs, and complementary local access technologies must interoperate in a manner that enhances continuity of service and provides fallback paths under adverse conditions, but this comes with complex challenges in terms of latency, mobility management, protocol translation, and orchestration across heterogeneous standards and operational domains.

### 4.4.1  Dynamic 3D RAN

Dynamic 3D RAN refers to the expansion of the RAN into vertical and multi-domain dimensions. As shown in Figure 4.3, this encompasses mainly aerial and space-based infrastructures, including UAVs, HAPSs, and satellite constellations (and even space-based infrastructures supporting lunar exploration or planetary habitats in longer-term visions), but also non-conventional terrestrial deployments.[7] A practical example of non-terrestrial deployments for resilience is small on-wheels BSs (Dressler, et al., 2019). For instance, first responders to natural disasters, such as national guards, may realise a mesh RAN among their trucks, working as diesel-powered BSs, upon entering the environment. This is similar to UAV/HAPS but a more natural extension to current disaster management protocols. Also, small-cell on-car BSs, such as autonomous shuttles, may help provide densification on demand or defence against active jammers with wireless backhaul to existing network infrastructure (or uplink to HAPS/satellite).

Today, many design and technology choices focus either on robustness (e.g., hardening satellite coverage) or on reactive challenge handling (e.g., UAVs/HAPSs deployed after disasters or during jamming), which is a too-narrow perspective. Indeed, a broader resilience view must also include challenge reaction and proactive adaptation, as discussed in Chapter 3. For example, UAVs and HAPSs can be repositioned not only reactively but also proactively as network demand shifts, while satellites may support reactive functions such as beam reallocation, adaptive steering, or (in the longer term) dynamic orbit adjustments.

---

[7] There are also underwater and underground systems, but these contribute less to wide-area resilience. Their relevance lies mainly in ensuring the robustness of critical terrestrial infrastructures they support (e.g., subsea backbones) and in maintaining local survivability in mission-critical environments.

The high dynamism of the 3D RAN architecture is both its greatest strength and its central challenge. Indeed, the harmonious coexistence and interplay of the domains requires standardised interfaces and protocols for each terrestrial, aerial, or orbital layer to announce its service availability, coordinate with others, and serve UEs seamlessly. This becomes especially complex when multiple stakeholders and jurisdictions are involved, demanding technical and regulatory alignment on access rights, SLAs, liability, and cost-sharing.

To prevent cascading failures across such a tightly interconnected architecture, containment strategies must be deployed at multiple levels. These include:

- Fail-safes, such as predefined fallback modes. For example, when a satellite link becomes unreliable, the system may revert to terrestrial or HAPS-based alternatives, or maintain minimal control-plane functionality via persistent, low-bandwidth backbones. Similarly, edge nodes can be pre-provisioned to operate autonomously when uplinks are unavailable. 3GPP Rel-18 already introduced NTN integration with basic graceful degradation procedures (e.g., fallback to direct LEO gateways if terrestrial midhaul collapses), but limited to a few well-defined cases (3GPP, 2023). To mature, future releases need to expand fallback handling to cover a broader range of TN–NTN interplays, and to standardise cross-domain signalling for autonomous failover.

- Isolation zones, which are logical or physical segmentation of network components to localise failures and prevent fault propagation across domains. A malfunction in a satellite cluster or UAV group, for instance, can be contained to prevent impact on TN or adjacent NTN segments. O-RAN specifications already provide partial support for this via network slicing and security domains, but these are primarily designed for steady-state operation. For resilience, isolation needs to be extended to dynamic and *ad-hoc* contexts.

- Dynamic reconfiguration, which entails real-time network adaptation and allows orchestration systems to reroute traffic, spin up BSs (vehicle-mounted, UAV-based), or shift loads adaptively in response to adverse events. From our discussions in Section 4.2, programmability and AI-native designs with intent-based management and real-time telemetry are crucial here.

Figure 4.3
Description: dynamic 3D RAN domains

*Figure 4.3: Dynamic 3D RAN domains.*

Finally, synchronised orchestration across domains is needed to avoid control-plane fragmentation. Here, TSN's multi-grandmaster architecture can be mirrored in the RAN using GNSS-synchronised grandmasters or boundary clocks across TN and NTN nodes.

### 4.4.2 Hybrid access networks

While dynamic 3D RAN focuses on vertical integration of TN and NTN, hybrid access networks emphasise horizontal integration of heterogeneous technologies. These are also crucial as backup paths and complementary fabrics, as discussed below.

- Wired infrastructures, such as Ethernet networks enhanced with TSN standards, such as TTEthernet and IEEE 802.1 TSN, are already foundational in industrial and critical applications. TSN provides ultra-precise time synchronisation, bounded latency, and fault tolerance, making it ideal for backhauling distributed RAN components, such as O-RAN's RUs, DUs, and CUs, in 6G. In fact, TSN's multi-grandmaster model could inspire distributed timing across heterogeneous nodes. Meanwhile, in the opposite direction, URLLC and eMBB capabilities in 5G/6G may also serve as backup paths in case of failures in the wired network.
- Local RATs, e.g., Wi-Fi, can serve as auxiliary paths for local traffic rerouting, especially within private or campus environments. Their lower deployment costs and fine-grained local control make them suitable for extending wireless coverage indoors or in rapidly deployed outdoor scenarios. Resilience can be realised and promoted with multi-RAT orchestration by diversifying frequency bands, protocols, and physical paths, reducing susceptibility to congestion, fading, or targeted jamming.
- Optical wireless communications, including visible light and infrared, can offer high directionality, immunity to RF interference, and large bandwidth. Hybrid radio-optical systems may dynamically shift traffic between radio and optical domains depending on interference or load conditions, increasing availability and resilience, and even leveraging existing infrastructure for sustainable communication.

The integration of these diverse technologies is non-trivial as it requires protocol translation, QoS alignment across domains, and synchronisation orchestration (Masaracchia, et al., 2023) (Reifert et al., 2024). For example, translating TSN's redundancy mechanism (IEEE 802.1CB FRER) into 6G requires mapping redundant TSN streams to 6G flows over the respective tunnelling protocols and distributed RAN configurations. This also requires specific Ethernet-based packet classifications (e.g., VLAN tags and priorities) to be supported, necessitating edge translator functions between the TSN and 6G domains, an area that is being actively researched. Meanwhile, while cellular networks define service differentiation via 5G QoS identifiers, TSN uses eight traffic classes, and Wi-Fi typically operates with four access categories; hence, harmonising these requires careful QoS mapping strategies to ensure consistent end-to-end latency guarantees, reliable packet delivery, and congestion control across the entire hybrid network. Beyond these, seamless end-to-end management across domains governed by different standards (e.g., 3GPP vs IEEE) remains a significant research challenge.

## 4.5 Architectural takeaways

Discussions throughout the chapter support the notion that resilience in 6G cannot be achieved by layering protective mechanisms onto fragile foundations. However, it must be engineered into the architecture from the outset. Modularity, disaggregation, redundancy, diversity, programmability, observability, AI-native control, and sustainability are the E2E design pillars for a truly resilient 6G.

Resilience requires hybrid control across multiple domains, balancing central intent with autonomous local loops. This does not simply mean loose coupling or redundancy in isolation, but rather coordinated mechanisms that ensure faults are absorbed locally without losing global coherence. Additionally, beyond ensuring availability guarantees for known faults, 6G architectures must also degrade gracefully under unforeseen conditions. This is facilitated by software-defined transport, intent-driven orchestration, and digital twin replicas, which can provide early warnings of critical events, automated recovery actions, and proactive explorations of "what-if" scenarios. The AI-native orchestration on top of this can only deliver its promise when paired with hybrid governance. This entails i) fast, localised decisions taken by AI agents; ii) human-in-the-loop oversight to manage exceptional or ambiguous conditions; and iii) architectural fallback modes that sustain minimal service when both automated and human control are compromised.

All in all, resilience must extend across the full continuum of edge, transport, compute, timing, and energy resources, while making their dependencies explicit and actionable. We emphasise that transport should not be treated as an external constraint but should be tightly integrated into this continuum. Taken together, these requirements and mechanisms transform resilience from an add-on feature into a defining property of the 6G architecture.

# 5 Technological enablers towards resilience

As future networks are expected to serve critical infrastructure, autonomous systems, and dynamic service environments, their ability to withstand, adapt to, and recover from disruptions becomes a foundational requirement. To meet these demands, technological innovation is not only a key enabler but also a strategic driver of systemic resilience.

This chapter examines how specific technology domains contribute to the development of resilient 6G systems. We categorise the enablers into three core areas: *networking techniques* focusing on architectural strategies such as connectivity diversity, virtualisation, decentralisation, and energy-aware design that enhance the resilience of the 6G network; *AI methods* investigating online optimisation in proactive fault management, autonomous reconfiguration, and agentic decision-making; and *security and trust designs* highlighting how cybersecurity, zero-trust frameworks, post-quantum cryptography, and trust management mechanisms underpin resilient operation in adversarial and uncertain conditions. Together, these technologies form a multi-layered foundation upon which resilient 6G capabilities can be built and sustained.

## 5.1 Networking techniques

## 5.2 AI methods and algorithms

AI methods and algorithms are foundational to enabling resilient, adaptive, and efficient 6G networks. By leveraging advances in predictive optimisation, autonomous decision-making, decentralised learning, and generative modelling, AI empowers networks to anticipate failures, dynamically reconfigure resources, and maintain robust operation under uncertainty. These intelligent capabilities transform network management from reactive to proactive, ensuring continuous service and optimised performance in complex and evolving wireless environments.

> Figure 5.2
> Description: The complementary role of different AI approaches from 5.3.1-5.3.3 in ensuring adaptive, predictive, and self-organising resilience.

*Figure 5.2 The complementary role of different AI approaches from 5.3.1-5.3.3 in ensuring adaptive, predictive, and self-organising resilience.*

### 5.2.1 Predictive and adaptive AI for resource management

Artificial intelligence–driven predictive optimisation and fault management are central to building resilience into 6G networks. Leveraging real-time data streams and machine learning models, these systems can anticipate and address potential failures before they occur, enabling proactive maintenance, minimising service disruptions, and sustaining performance under

6G Resilience White Paper – DRAFT                                                                                                                40 fluctuating conditions. By combining predictive analytics with adaptive control, they allow dynamic resource allocation, rapid mitigation of failures, and continual adaptation to evolving network states. Online learning further enhances robustness by updating models in real time, ensuring that decision-making remains effective even in non-stationary or adversarial environments. This evolution from reactive fault handling to proactive, resilient-by-design management is key to meeting the operational demands of next-generation infrastructure.

In the context of resource optimisation under uncertainty, maintaining reliable service delivery in the presence of unpredictable dynamics is a critical challenge. Uncertainty in wireless systems arises from multiple sources, including imperfect channel state information caused by noise, outdated measurements, or limited feedback, as well as the non-deterministic computational demands of tasks offloaded to edge or cloud servers. These factors can degrade link quality and destabilise performance if not addressed systematically. Conventional deterministic allocation strategies are often inadequate, as they fail to safeguard against worst-case conditions. AI-based robust optimisation methods address this gap by explicitly incorporating uncertainty into the decision process. For instance, deep neural networks (DNNs) can be trained using uncertainty injection, where sampled variations in channel quality and computational demand are used to generate resource allocation decisions, such as transmit power, bandwidth assignment, and computation offloading, which are then evaluated across simulated adverse scenarios. By optimising against a percentile-based loss reflecting worst-case performance, such models develop the capacity to maintain service quality under unpredictable conditions. Modular DNN designs, which train separate models for different uncertainty parameters, can simplify computation but may underperform compared to joint optimisation approaches that capture interdependencies between variables.

Adaptation to evolving operational states requires learning paradigms that can update decision-making policies in real-time. Online learning methods (Shalev-Shwartz, 2012) support incremental model refinement as new data becomes available, thereby avoiding the rigidity of fixed, offline-trained models. This continuous adaptation is particularly advantageous in non-stationary environments, where channel conditions, traffic patterns, or adversarial threats change unpredictably. Rollout algorithms with multi-step lookahead (Silver, et al., 2017) and (Bertsekas, 2021), grounded in adaptive and approximate dynamic programming (Bertsekas, 2022), provide a principled framework for sequential decision-making under uncertainty. By simulating future trajectories and iteratively refining policies based on observed outcomes, rollout-based controllers can anticipate the long-term impact of present actions and adjust accordingly. This ability to integrate predictive foresight with real-time feedback makes them highly effective for sustaining resilience in complex, high-dimensional 6G environments.

### 5.2.2 Decentralised and autonomous AI for network adaptation

Conventional centralised management approaches can be too slow or communication-intensive to meet the demands of dynamic and real-time reconfigurations and adaptations, especially in large-scale, heterogeneous, and mission-critical environments. Artificial intelligence (AI) offers a pathway to achieve distributed, autonomous control through the integration of agent-based intelligence, federated learning, and self-optimising network (SON) capabilities. In this context, decentralised AI frameworks enable localised decision-making by equipping network elements, such as base stations, edge nodes, or unmanned aerial vehicles, with embedded learning agents. These agents can sense local conditions, predict upcoming requirements, and coordinate with peers to enact reconfiguration actions without relying on a central controller for constant direction. Such an approach not only reduces latency and communication overhead but also enhances robustness against single points of failure (Ale, et al., 2025).

Federated learning techniques play a crucial role in this paradigm by allowing models to be collaboratively trained across distributed agents while keeping data local. This preserves privacy (Siriwardhana, et al., 2024), reduces backhaul traffic, and allows agents to benefit from global knowledge while maintaining adaptation to local contexts. In parallel, agentic AI systems can incorporate intent-based policies and reinforcement learning to autonomously allocate resources, re-route traffic, or adjust parameters in response to evolving conditions. The combination of agent-based control and federated intelligence can support a wide range of resilience-oriented functions, including proactive fault detection, predictive maintenance, and self-healing. Moreover, by leveraging hierarchical or peer-to-peer coordination strategies, the network can respond to disruptions, such as node failures, congestion events, or interference spikes, through rapid and targeted reconfiguration (Zhu, et al., 2022). These capabilities will be essential in enabling the autonomous, adaptive, and resilient behaviour required for beyond-5G and 6G systems.

### 5.2.3 Generative and model-mased AI for planning and simulation

Generative models enable the simulation of diverse failure and recovery scenarios, allowing networks to prepare proactively for rare or extreme events that may otherwise be overlooked. Meanwhile, model-based learning integrates domain knowledge and physical constraints to improve the reliability and interpretability of AI systems, particularly in environments where data availability is limited. Together, these methods support robust and foresighted control strategies, empowering networks to anticipate and effectively navigate disruptions.

Generative AI (GenAI) significantly advances digital twin systems, which are virtual replicas of physical network components, by facilitating the real-time simulation of fault scenarios, anomaly detection, and predictive maintenance. These models can produce synthetic data that mimics rare or complex fault conditions, enhancing the network's resilience and fault preparedness. For instance, GenAI-powered digital twins can proactively identify failure patterns and recommend optimal recovery strategies, thereby minimising downtime and bolstering system reliability (Curic & van Maastricht, 2024). In telecom operations, generative AI aids automated fault detection, network resource optimisation, and prescriptive analytics by forecasting traffic congestion, simulating recovery actions through network twins, and generating scripts for automated fault resolution. These capabilities span the layers of data, analytics, and automation, enabling GenAI to unify vendor-specific data, predict future network states, and simulate optimal recovery actions prior to deployment. Furthermore, by combining generative AI with simulation frameworks and retrieval-augmented generation (RAG), operators can evaluate multiple recovery paths within virtual environments, ensuring the selection of the most effective strategy without disrupting live systems. This approach supports essential functions, such as graceful degradation, state replication, and automated rollback, which preserve operational integrity in the event of faults. The integration of generative AI with agentic AI workflows and edge computing further enhances real-time, decentralised fault management, a capability crucial for achieving ultra-reliable, low-latency communications in 6G.

Model-based machine learning offers complementary advantages in large-scale, dense wireless deployments, where AI-based signal processing is expected to play a central role. While AI has influenced 5G systems, its use has typically relied on abundant data and powerful computing resources. Edge devices, however, often face significant challenges, including limited training data, unreliable inference inputs, and constrained power budgets that complicate stable AI operation, especially when online learning is required. Model-based ML addresses these challenges by incorporating established physical and mathematical models of wireless channels, noise, and interference into the learning process (Zappone, et al., 2019). This produces solutions that are both more robust and interpretable, especially in scenarios characterised by limited data, resource constraints, or rapidly varying conditions. Techniques such as deep unfolding, which map traditional iterative algorithms into neural network layers, enable efficient, low-complexity implementations that retain the interpretability and stability of classical methods (Nguyen, et al., 2021). Moreover, model-based ML inherently exhibits greater resilience to data distribution shifts, such as sudden channel fluctuations or hardware impairments, since it leverages the optimisation structure of algorithms rather than relying solely on patterns learned from data. This approach also adapts well to resource constraints: when learning is infeasible, model-based ML architectures can seamlessly revert to conventional optimisation procedures using algorithmic rules without requiring training.

## 5.3 Security & trust designs

Security and trust are fundamental prerequisites for the resilience of 6G networks, which will operate in highly heterogeneous, dynamic, and adversarial environments. The increasing reliance on cloud-native infrastructures, AI-native network functions, and pervasive connectivity across critical domains such as industrial automation, healthcare, and transportation amplifies the attack surface and exposes networks to novel vulnerabilities. Unlike previous generations, 6G must adopt security and trust designs that are not merely reactive but proactively adaptive, embedding resilience into the system's architectural fabric. This requires moving beyond perimeter-based protections to embrace Zero-Trust Architectures (ZTAs), integrating Post-Quantum Cryptography (PQC) to counter emerging quantum-enabled threats, and developing trustworthy identity management and micro-segmentation mechanisms tailored for cloud-native services. Furthermore, the convergence of AI-driven automation with advanced cryptographic and trust technologies calls for a holistic security framework that ensures interoperability, energy efficiency, and ultra-low latency without compromising robustness. In this context, 6G security and trust designs must evolve from isolated safeguards to cohesive, systemic strategies that continuously adapt to evolving adversarial dynamics, guaranteeing integrity, confidentiality, and availability across the full lifecycle of network operations.

> Figure 5.3.
> Description: The schematics of zero-trust workflow showing identity verification, micro-segmentation, and AI-based intent management towards network resilience

*Figure 5.3 The schematics of zero-trust workflow showing identity verification, micro-segmentation, and AI-based intent management towards network resilience.*

### 5.3.1 Resilience through proactive threat prevention and detection

Traditional reactive security mechanisms are inadequate for addressing the dynamic, heterogeneous, and high-stakes nature of 6G networks, where resilience requires proactive threat prevention and detection strategies. Such approaches are crucial for ensuring service continuity, preserving data integrity, and maintaining trust in latency-sensitive and mission-critical applications, such as autonomous transportation, remote healthcare, and smart grids. Moving Target Defence (MTD) has emerged as a key enabler of proactive resilience by continuously altering the attack surface through mechanisms such as dynamic resource shuffling, software diversity, and functional redundancy, thereby inducing operational uncertainty that disrupts adversarial reconnaissance and complicates attack execution (Javadpour, et al., 2024). Beyond strengthening conventional security, MTD also enhances the robustness of AI/ML systems that are vulnerable to data poisoning, evasion, and model inversion attacks, by dynamically modifying models, features, or parameters to mitigate adversarial risks (Motalleb, et al., 2025). Conversely, the integration of AI/ML augments MTD by enabling real-time threat detection, automated analysis, and adaptive orchestration of defensive strategies, forming a self-adaptive framework vital for resilient 6G networks (Kianpisheh, et al., 2024). Complementary to MTD, intrusion detection and prevention systems provide another critical layer of defence by continuously monitoring network activities and identifying anomalies through model-based or data-driven approaches. Evolving from traditional signature-, heuristic-, or behaviour-based methods to advanced AI/ML-based systems, these mechanisms leverage supervised, unsupervised, and reinforcement learning to detect and predict threats such as jamming or spoofing, thereby ensuring network resilience against evolving attack vectors.

Current research on proactive threat prevention has essentially treated MTD and intrusion detection as separate paradigms, with limited integration into the broader AI-driven orchestration frameworks required for 6G. Hence, there is a need for advancements by tightly coupling MTD with AI/ML-enhanced intrusion detection and prevention, forming a unified, adaptive, and context-aware defence architecture. Such integration moves beyond static or reactive defence postures toward a closed-loop system capable of anticipating, simulating, and counteracting evolving adversarial strategies in near real time. Towards this, novel mechanisms that can dynamically fuse multi-source threat intelligence with network state information to enable predictive MTD reconfigurations that pre-emptively reduce attack success probabilities, embed adversarial robustness aspects directly into the learning pipelines to ensure continuity of learning-driven functions under attacks, and leverage reinforcement learning and cross-domain knowledge transfer to detect novel or stealthy attack patterns without prior exposure are to be developed. Together, these innovations can establish a new foundation for proactive, resilient security in 6G, where prevention, detection, and adaptation are jointly optimised to deliver trustworthiness in highly dynamic and mission-critical environments.

### 5.3.2 Zero-trust architectures for network integrity

ZTAs offer a foundational paradigm for ensuring network integrity in 6G systems by assuming that no entity, whether internal or external, is inherently trustworthy. A central principle of ZTA is micro-segmentation, which strengthens resilience by creating micro-perimeters around critical resources and enforcing fine-grained, context-specific security policies. This reduces the attack surface and prevents adversaries from moving laterally within the network, while providing enhanced visibility into traffic flows that support proactive anomaly detection and rapid containment of breaches. Emerging inter-service communication technologies such as Service Mesh and Container Network Interface (CNI) have been identified as promising enablers of truly end-to-end Zero-Trust micro-segmentation for cloud-native 6G infrastructures (Benzaïd, et al., 2025). With built-in mechanisms such as mutual Transport Layer Security (TLS) authentication, fine-grained access control, and service-level traffic observability, these technologies enable dynamic, policy-driven enforcement of least-privilege principles, thereby addressing both external and insider threats. Moreover, the integration of AI, particularly through Large Language Models (LLMs), enhances intent-based management of micro-segmentation policies, enabling their automated derivation, enforcement, and adaptive refinement throughout the policy lifecycle.

Complementing micro-segmentation, identity management, and trust anchors are critical for ensuring that only authorised entities gain access to networks, functions, and services. Robust authentication and end-to-end encryption mechanisms mitigate impersonation, interception, and replay attacks across untrusted networks, securing communications in diverse and mission-critical applications. Nevertheless, existing trust establishment mechanisms, including blockchain-based approaches (Hewa, et al., 2022), remain vulnerable to adversarial strategies such as collusion, highlighting the necessity of continuous security monitoring, rapid incident response, and adaptive trust management. The emergence of AI-native functions further underscores the importance of reinforcing trust mechanisms, as adversarial AI-based attacks present new risks to the integrity and resilience of 6G systems.

Towards enabling a dynamic and intelligent Zero-Trust ecosystem for 6G, the use of AI-driven intent-based mechanisms, powered by LLMs, to autonomously generate, validate, and adapt micro-segmentation policies in real time is needed. Unlike conventional methods that rely on preconfigured policies or rigid rule sets, the integration of AI with Service Mesh and CNI technologies enables fine-grained, context-aware enforcement that continuously evolves with the threat landscape (Benzaïd, et al., 2025). Additionally, vulnerabilities of blockchain-based trust anchors, such as collusion attacks (Hewa, et al., 2022), are addressed by coupling distributed trust mechanisms with adaptive monitoring and rapid incident response, creating a more resilient trust fabric. These could yield a Zero-Trust framework that does not merely react to adversarial activity but anticipates and adapts to it, ensuring strong identity assurance, minimised attack surfaces, and integrity of AI-native functions under adversarial pressure. Such proactive and self-adaptive trust architectures represent a significant step beyond existing Zero-Trust designs, which are instrumental for secure and resilient 6G networks.

### 5.3.3 Post-quantum cryptography

Ensuring resilience against adversaries equipped with quantum computational power is becoming an indispensable requirement for 6G networks. Classical cryptographic algorithms for symmetric and asymmetric key encryption, which safeguard privacy and integrity today, rely on hard mathematical problems such as the Discrete Logarithmic Problem (DLP). However, such schemes are rendered vulnerable in the quantum era, as algorithms like Shor's algorithm (Shor, 1994) can efficiently solve these problems in polynomial time. Recognising this, the National Institute of Standards and Technology (NIST) has accelerated the development of post-quantum cryptographic standards, underscoring the urgency of transitioning to cryptographic primitives resilient to quantum attacks. Preparing communication infrastructures for this post-quantum era requires significant scientific and engineering contributions to ensure seamless interoperability and compatibility across heterogeneous systems. Furthermore, the unique threat scenario of "collect now, decrypt later" accentuates the importance of adopting PQC solutions early, as adversaries may store encrypted communications now with the intent of breaking them once quantum capabilities become available.

In the context of 6G, PQC is envisioned as a foundational security enabler against quantum-aided adversaries and increasingly complex threat scenarios. Standardization bodies such as 3GPP are already specifying guidelines for quantum key distribution, establishing the groundwork for quantum-safe security in the early phases of 6G deployment. The diverse application domains of 6G, including mission-critical sectors like Industrial IoT, demand cross-platform compatibility and interoperability, which must be preserved throughout the cryptographic transition. At the same time, maintaining the ambitious design targets of 6G such as energy efficiency, ultra-low latency, and massive scalability, poses further challenges, since PQC algorithms often introduce larger key sizes and higher computational overhead.

The advancements in the integration of PQC into the broader 6G ecosystem in a holistic manner are essential to jointly optimize the aspects of security, performance, and interoperability. Beyond focusing solely on cryptographic strength, novel methods should consider lightweight and energy-efficient PQC implementations that can be deployed across resource-constrained devices and latency-sensitive environments. Furthermore, the combination of PQC with emerging 6G-native paradigms such as zero-trust security and AI-driven orchestration enables adaptive, context-aware resilience that evolves with the threat landscape. By ensuring interoperability across heterogeneous platforms while minimising performance penalties, this approach could be a promising research avenue towards achieving sustainable and quantum-resilient 6G infrastructures that extend beyond the scope of current PQC standardisation initiatives.

## 5.4 Toward comprehensive resilience in 6G

Isolated mechanisms or localised safeguards cannot guarantee the resilience of 6G networks. Instead, it must emerge from the seamless integration of distributed architectures, AI-native algorithms, and robust security frameworks into a unified and adaptive ecosystem. The discussions throughout this chapter emphasise the need to consider resilience holistically, encompassing networked infrastructures, intelligent resource management, security and trust, and the integration of autonomy and adaptability at every level. The interplay between distributed and decentralised designs (Section 5.2), advanced AI methods and algorithms (Section 5.3), and comprehensive security and trust architectures (Section 5.4) demonstrates that resilience is not a static property but a dynamic, evolving capability that must be continuously reinforced as networks grow in complexity and exposure.

From the systems perspective, several implications for further study and research questions were found. How to enable cyber-physical integration of physical assets and data while ensuring interoperability across diverse hardware, vendors, and platforms? How can we account for component and network-level evolutions that result in different levels of resilience between components and elements? How to organise the required effort needed with semiconductor technologies and the digital supply chain? How to define cross-layer and cross-infrastructure KPIs for a complex 6G system, e.g., to consider time to detect, remediate, short-term recovery, and long-term adaptation? How to define and potentially standardise multiple operation modes, resilience metrics, and evaluation frameworks in a single system level and across interdependent systems? How to assess the overall impact on applications' service levels and availability while their resilience requirements differ.

Looking forward, several avenues for future research emerge. First, a deeper exploration is needed into how distributed intelligence and decentralised architecture can be co-designed with AI-driven prediction, adaptation, and learning mechanisms to ensure both robustness and efficiency under adverse conditions. Second, new opportunities arise from the convergence of generative and model-based AI with security frameworks, enabling the proactive detection, simulation, and mitigation of threats in real-time. Third, integrating PQC, Zero-Trust principles, and intent-driven security into cloud-native and heterogeneous environments requires significant advances in cross-layer design, interoperability, and energy-efficient implementations. Ultimately, comprehensive resilience in 6G will necessitate rigorous simulation, validation, and large-scale experimentation to assess the effectiveness of these approaches in realistic settings.

Ultimately, resilience in 6G will be defined by the network's ability to anticipate, withstand, adapt to, and recover from disruptions without compromising the stringent requirements of latency, reliability, scalability, and sustainability. Achieving this vision calls for collaborative, interdisciplinary research that bridges communication theory, AI, cybersecurity, and system design, ensuring that 6G becomes a secure, adaptive, and trustworthy foundation for the digital ecosystems of the future.

# 6 Techno-Economics of Resilient 6G System

A techno-economic examination of 6G resilience reveals a diverse range of resilience-focused business model designs in 6G that use technological features to create economic outputs, impacting sectors. The planned open, resilient 6G platform can support a developer community to innovate technologies, services, and business models, increasing network value and opportunities.

## 6.1 Resilience forces - trends and uncertainties

The forthcoming generation of mobile communications, 6G, is anticipated to emerge as a *general-purpose technology* (GPT) platform (Teece, 2018) that is ubiquitous and has the potential for continuous technical improvement. 6G is expected to enable innovation complementarities both up and down the sectoral value streams, as is the case with the preceding generation, but triggering spillover effects across the various industry sectors utilising 6G. The definition of *resilience* here, rooted in social-ecological thinking, extends to cover "*the capacity to deal with change and continue to develop*," (Stockholm Resilience Centre, 2020). In the context of wireless systems, a resilient system is one that "*is prepared to face challenges, withstand them, and prevent most from causing performance degradation. It can also absorb the impact of significant challenges, ensuring essential functionalities or a minimum service level. Moreover, it can recover, adapt, and evolve based on the experiences learned during this process*" (Khaloopour, et al., 2024).

To fully realise the cross-sectoral potential of resilient 6G, a reassessment of how a diversity of network resources, services, and applications are created, delivered, shared, and consumed across various segments of a resilient digitised society is essential. Technological advancements are poised to reshape business models and alter the roles of established stakeholders within the resulting ecosystems. These changes will also lower barriers to entry for emerging actors, such as digital service operators, platform providers, and resource brokers (Yrjölä, et al., 2023). The open system architecture and the ability to control networks within applications will be particularly transformative in 6G for a diverse range of application developers (Yrjölä, S., 2024).

The global security and trustworthy environment is an ongoing negotiation, characterised by increasing social instability, heightened technology competition, geopolitical conflicts, and a heightened risk of future armed and malicious actors' engagements. Societies in Europe rely heavily on critical infrastructures to maintain the normal operations of social systems, which set increasing demands for inclusivity, sustainability, resilience, and transparency in technological innovations (Matinmikko-Blue, et al., 2020). The foundation for developing a resilient future 6G starts with acknowledging the actions taken to safeguard national sovereignty and addressing the importance of finding ways to respond to the global grand challenges of sustainability through transformative and goal-oriented innovation policies, as well as anticipatory and flexible regulation (Ahokangas, et al., 2023). The national emergency agencies have prioritised critical vertical sectors, including energy, food and water supply, healthcare, transport and supply chains, digital security, finance, and communications. Military customers' and public authorities' needs for deployment of, and integration with, public mobile networks will create increasing demands for resilience in 6G and change the business models applied in 6G. A robust and secure digital infrastructure is considered crucial for a nation's competitiveness, sovereignty, and resilience (NATO, 2023). The telecommunications infrastructure can be considered a prime target in overt conflict, as evidenced by the situation in Ukraine. Consequently, public safety and military budgets are escalating globally, potentially diverting resources from investments in sustainability and digital initiatives and influencing technological choices.

The trends identified and summarised in **Error! Reference source not found.** create new opportunities and funding streams, impacting the trajectory of technological development. 'The market' is generally averse to conflict, as evidenced by negative responses to war, erosion of sovereignty, climate change, and widespread cyberattacks. Furthermore, businesses are not inherently driven to prioritise resilience in support of public goods, such as strategic autonomy, economic security, and national security (Timmers, P., 2024). In the realm of public policy, resilience is a fundamental component of economic security, which in turn is an integral aspect of strategic autonomy. *Economic security* encompasses not only the promotion of resilience but also the safeguarding of economic assets and the fostering of trust in international economic relations associated with supply chains, physical and cyber security of critical infrastructure, technology security and technology leakage, and weaponisation of economic dependencies or economic coercion (European Commission, 2023). *Strategic autonomy*, in turn, serves as the cornerstone of safeguarding national sovereignty. It encompasses the development and maintenance of capabilities (knowledge and skills) and capacities (production, manufacturing, and services), along with control over these assets, enabling a nation to independently determine and execute its future trajectory in the domains of economy, society, and democracy. This understanding has led to a growing trend among governments towards the adoption of *strategic capitalism*. This involves active government intervention in the economy to boost specific industries and improve national competitiveness and resilience through industrial policy, regulation, R&D investment, and other targeted initiatives (Spence, 2011).

Rapid technological growth, global activity reach, and enforcement issues pose challenges for regulators, possibly leading to gaps in resilience regulation. Inconsistent international digital regulations can enable regulatory arbitrage and a 'race to the bottom,' where jurisdictions lower standards and undermine broader policy goals (Sarliève, P. et al., 2025). Resilience relates to sustainability, defined as actions that do not compromise future options in economic, societal, and environmental contexts (Elkington & Rowlands, 1999), emphasising efficient resource use for survival amid change. While sustainability implies resilience, the reverse isn't always true due to trade-offs, such as redundancies. In edge-cloud architectures, decentralisation improves fault tolerance but increases distributed components needing power and maintenance. Resource-saving efforts may also limit rapid disruption responses. Extending coverage to rural areas is key for digital inclusion and resilience but can raise energy costs and infrastructure needs in low-density regions.

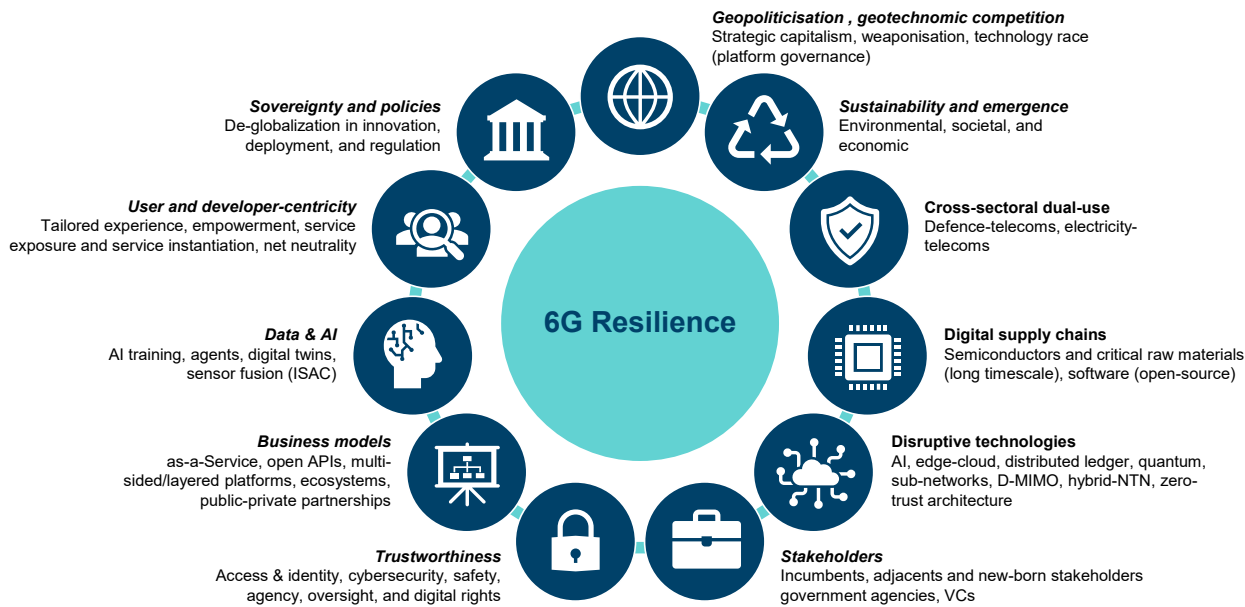

*Figure 6.1 Trends and uncertainties impacting 6G resilience*

## 6.2 Techno-economic lenses to explore a resilient 6G system

Resilience is a nascent area in mobile techno-economic research. Efforts haven't fully addressed the resilience business models' systemic and platform-based nature in mobile networks. This section outlines frameworks for evaluating resilient 6G systems and maintaining healthy, fair markets.

*The business model (BM)* offers a framework that transforms technological features and potential into economic outputs via customers and markets (Chesbrough & Rosenbloom, 2002). Value creation, central to BMs and strategic management, reflects an organisation's operational logic, especially in digital societies. Key value drivers include transaction efficiency, complementarities, customer lock-in, and innovation (Amit & Zott, 2001). This study used the *action-based BM framework*, which explicitly considers stakeholders and actions as a proxy for exploring and exploiting resilience business opportunities. It builds on opportunity exploration and exploitation, linking partners and customers through value co-creation, sharing, and co-capture. Companies' connected business models underpin ecosystem thinking. The focus on future change at the technological, BM, or ecosystem levels guides business model innovation (BMI) (Foss & Saebi, 2017) (Ahokangas, et al., 2022).

We adopted the *5C BM typology* (Yrjölä, S., 2024), based on the 4C model (Wirtz, et al., 2010), to identify and categorise the key technological system components of resilience BMs in the emerging 6G environments and platforms. The five prototypical models, each with varying value propositions and revenue models, are depicted in **Error! Reference source not found.**:

1. *Connection* model offers virtual and/or physical network infrastructure and related services needed to exchange information and users' participation.
2. C*omputing* model offers virtual and/or physical computing infrastructure and related services to other layers, providing the prerequisites for data, AI algorithms, and cloud business models.
3. C*ontent* model collects, selects, compiles, distributes, and/or presents various types of data.
4. C*ontext* model sorts and/or aggregates available data and provides structure and navigation for users to increase transparency and situational awareness while reducing complexity.
5. C*ommerce* model that initiates, negotiates, and/or fulfils online transactions, and enables low transaction costs for buyers and sellers of goods, data, and services.

The BM design for resiliency analysis focuses on how a company structures its interdependent activities to determine the focus, locus, and modus of its business (Onetti, et al., 2012), which transcends the focal company and spans its boundaries (Zott & Amit, 2010).

- *Focus* of business decisions involves allocating resources to core value activities, determining where to invest or divest, and defining the relevance of activities and the value chain span.
- *Locus* decisions specify where resources and activities are located geographically or within clusters.
- *Modus* shapes how a company operates, managing activities through internal organisation, outsourcing, and collaboration, considering capital, technology, operations, and skills.

## 6.3 Resilient business model design in 6G system

### 6.3.1 Technological and architectural enablers

The key technological and architectural enablers can be interpreted as business model antecedents, and their components are categorised based on their main contributions to the 3R-perspectives of system resilience and the 5C-business model layers, as summarised in **Error! Reference source not found.**.

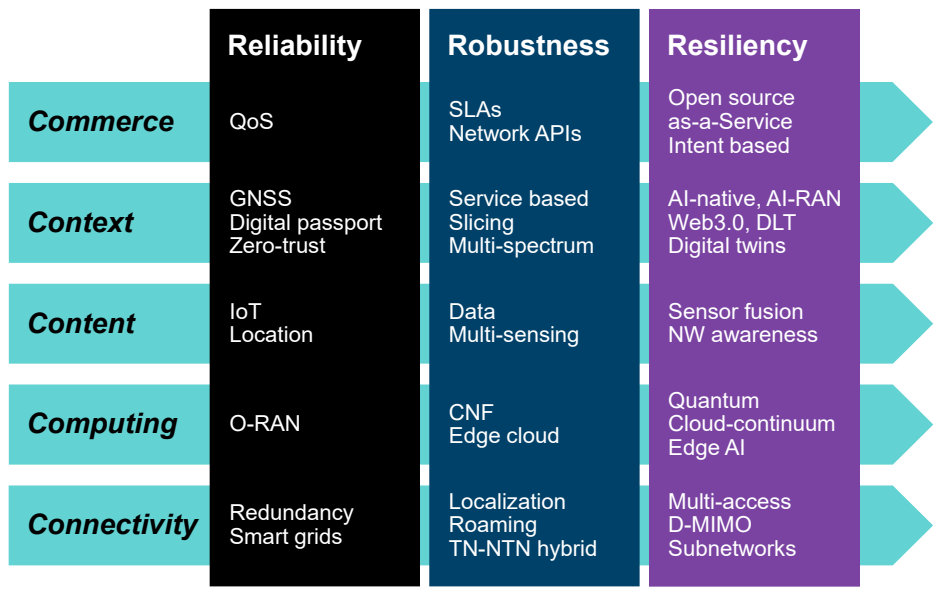

*Figure 6.2 Key enabling technologies as business model antecedents for reliable, robust, and resilient 6G business models.*

## 6.4 Business model design patterns for a resilient 6G system

We present nine business model patterns for resilient 6G systems based on stakeholders like mobile network operators (MNOs), vendors, industry groups, and agencies. These patterns draw from environmental, social, and governance (ESG) reports, sustainability reports, corporate social responsibility (CSR) reports, and white papers on 6G visions. The analysis categorised models for further study (Lüdeke-Freund, et al., 2018). Finally, business model pattern groups were assessed and characterised based on their value drivers (Onetti, et al., 2012) and platform archetypes (Springer, et al., 2025).

**1. Differentiated pricing and revenue**

The first business model design pattern group focuses on new revenue and pricing methods, such as differential pricing, social freemium, innovative financing, and service-driven models. SLA policies can be combined with risk-based service logic to achieve cross-platform effects with related industries. Co-pricing, revenue sharing, and purchasing adjustments, with the support of governments and NGOs, help societal resilience.

**2. Financing**

Fragmented funding and investment structures hinder scalable resilience solutions. Patterns show different ways of financing network reliability, robustness, and resilience upgrades. Critical infrastructure financing often uses public-private partnership (PPP) 'strategic capitalism' models with strong lock-in effects. The European Commission's upcoming Digital Networks Act (DNA) aims to promote resilient, secure digital networks in the EU, focusing on investment and regulation. Co-funding occurs across sectors like public safety, energy, cloud, and healthcare, especially with energy sustainability and smart grid initiatives through energy-aware resource management and green redundancy. New defence VC and sovereignty funds can also support dual-use technology innovations.

**3. Resilience design**

The pattern group recommends ways to boost 3R-resilience in activities, processes, products, and system design for market entry. Data, AI, deep tech, enabling tech, redundancy, subnetworks, and cybersecurity are common across stakeholders. 6G platform innovations need specialisation or customisation with domain-specific functions for complex use cases. Resilient designs must integrate physical assets with digital infrastructure to improve resilience. For example, zero-energy nodes and energy networking help at the device and edge levels, but larger-scale resilience requires coordination between telecom and energy systems. Early testing of components is essential for resilient 6G design. Standardisation innovations related to this are discussed in Section 6.5.3.

**4. Digital supply chain**

Companies and governments aim to address supply chain efficiency and digital platforms' centralised power (Srnicek, N., 2017) via supplier diversification, re-shoring, investing in redundancy, 'just-in-case' next to 'just-in-time', monitoring, and international coordination. Trade-offs balancing resilience, performance, economic efficiency, and sustainability is complex, especially in the short term. Supply-side opportunities linked to O-RAN, AI-RAN, open-source, digital product passports (DPP), circular economy, and cross-border cooperation rely on platform interoperability. The DPP aids resilience in digital supply chain design for products, apps, and services. Global policy tools include public funding, tax incentives, and fast-track approvals, which support strategic focus and policy teams. The EU Critical Raw Materials Act (CRMA) (European Union, 2024) aims to ensure a secure, resilient, and sustainable supply of critical raw materials (CRMs) for the EU's green and digital transitions and defence sectors. Moreover, the EU Chips Act exemplifies a comprehensive policy response to the geopolitical significance of semiconductor chips. This three-pillar program focuses on: (1) building technological capacity; (2) investing in production capacity; and (3) managing semiconductor supply crises.

**5. Cooperative**

This group examines sharing economy-based business models using platforms to match supply and demand through network effects. Shared resources and services include networks, clouds, data, AI, security, edge, terminal, and government assets, with dashboards providing real-time information on network configurations. Examples of operator collaboration in infrastructure sharing reduce redundant investments and environmental impact via resource interdependencies. The shift to distributed computing enhances resilience by processing critical data near the point of need, such as at edge nodes in RAN, hospitals, or substations. Edge computing and AI utilise local data to enable real-time decision-making, improve privacy, secure data, and increase efficiency. Application platforms benefit infrastructure providers, developers, and service companies. ISAC supports situational awareness, safety, and new services like personalised healthcare, predictive maintenance, environmental monitoring, and smart city solutions, with opportunities in smart grids and disaster relief networks.

**6. Mobile access provisioning**

This pattern group provides mobile access for sharing information and encouraging user participation across domains, boosting market reach and economies of scale. BMs support the development of 6G for the resilience market by offering solutions that enhance scalability, replicability, and sustainability of lock-in business models. Cloud computing and AI, via satellite connections, enable seamless transfer of operations during terrestrial network disruptions, boosting resilience and reducing dependence on local infrastructure for business continuity.

Smart energy grids with distributed sources and storage require strong communications for real-time data and control, improving grid resilience. The reliance on variable renewable energy increases the need for smart control and reliable telecommunications, especially as electrical systems become more complex, which also raises cybersecurity risks that demand better protection and real-time threat detection.

Digital inclusion can support remote surgery, telemedicine, and patient monitoring, expanding healthcare access and emergency response. Opportunities arise for medical device makers, healthcare providers, and telemedicine firms to ensure continuity of care during crises, strengthening community resilience. Resilient 6G networks for transportation can enable autonomous vehicles, connected infrastructure, and advanced traffic management.

**7. Context awareness**

Context awareness BM uses telco sensor fusion to recognise and adapt to the environment, building on novelty. It creates new markets from real-time data from JCAS and IoT sensors, including ambient sensors. Cross-side network effect involves massive twinning, which sorts and aggregates data, providing governance and structure to increase transparency and reduce complexity. Valuable data for AI training, compliant with laws like the AI Act, enhances sustainability. Spectrum anomaly detection finds unusual radio signals to ensure security. Real-time data orchestration- collecting, storing, analysing, and interpreting- combined with AI digital twinning, fosters innovative product development.

**8. Service and performance**

This pattern group creates a resilience market by dematerialising physical products into user-centric functions, services, and results. Federated APIs for mobile networks allow developers to access extended capabilities like digital twins and models, fostering value creation. Platform provisioning generates network effects by enabling operators to leverage network capabilities. Combined with transaction platforms, it enhances technological complementarities, applications, and resilience business models for existing and new stakeholders. Resilience-based services, unlike traditional SLA optimisation, could derive value from sovereignty control and dual-use infrastructure, supporting a service-oriented approach.

**9. Social mission**

The business model leverages complementarities to promote societal sustainability, emphasising user trust, safety, digital inclusivity, personal freedom, and cultural connection. These elements ensure that advancements like 6G foster social resilience (SNS JU, 2025). Sharing economy business models support socio-economic empowerment, creating markets and cross-side network effects. Decentralised autonomous platforms (DAPs) enable local survivability, prosumerism, and trust-based systems, giving individuals and communities more control over digital data, technology, and societal issues. Climate resilience initiatives leverage 6G to develop strategies that mitigate the impacts of climate change, such as drought resistance and phenological monitoring, ensuring productivity amid environmental changes. Total defence 6G networks aim for seamless communication across RAN, NTN, and all-photonic networks.

*Table 6.1 The identified business model pattern groups, value creation patterns, and key activities in a resilient 6G system.*

| BM pattern groups | BM patterns | FOCUS (value proposition) | LOCUS (industrial cluster) | MODUS (operating model) |
|---|---|---|---|---|
| **1. Differentiated pricing and revenue** | Service-driven logic<br>Cross-platform network effects<br>Co-pricing and revenue sharing<br>Purchasing power | Novelty<br>Adaptability<br>SLA tailoring<br>Sovereignty/Trust-as-a-Service<br>Cybersecurity-aaS<br>Risk level-based<br>Private networks | Incumbent platforms<br>Critical infrastructures<br>Industry 4.0<br>Quantum<br>Web3.0, DLT<br>Defence<br>Communities | Consortium platforms<br>Collaborative governance<br>Partnership with verticals<br>NGOs |
| **2. Financing** | Public-private partnership funding<br>Cross-domain co-funding<br>Venture capital | Lock-in<br>Adaptability<br>RDI funding<br>Deployment subsidies<br>Regulatory incentives | Government agencies<br>Defence funds<br>Public safety<br>VCs | Public-private partnership<br>Regulators, localisation<br>Alliances<br>Venture capital |
| **3. Resilience-design** | Market penetration<br>Cross-platform network effects<br>Deep tech innovations<br>Connection, computing | Efficiency<br>Tangible assets, redundancies<br>Test service<br>Key vulnerabilities<br>Situational awareness<br>Adaptive capacity<br>Inclusive transparency | Incumbent / Adjacent born platform<br>Telecommunications<br>Research labs<br>Cyber security<br>Mission-critical verticals | Consortium platforms<br>Tethered digital platforms<br>Planning and controlling<br>Standardization (SDOs)<br>IPRs |
| **4. Digital supply chain** | Scale economics<br>Connection, computing, content, context | Efficiency<br>Adaptability<br>Complementarities<br>Open source<br>O-RAN and AI-RAN<br>Digital product passport (SW BOM) | Incumbent platforms<br>Semiconductors<br>SW<br>Data and AI<br>National, regional, and global regulation | Decentralised autonomous platform<br>Collaborative governance<br>Standardization<br>Cross-border<br>Policy & regulation |
| **5. Co-operative co-owners & co-managers** | Market segmentation<br>Cross-platform network effects<br>Connection, computing, content, context | Complementarities<br>Adaptability<br>Infrastructure and resource sharing<br>API management<br>AI as a Service<br>Computing platform<br>Energy production and storage<br>System integration | Born and adjacent platforms<br>Cloud platforms<br>AI platforms<br>Microgrids<br>Dual-use civil-military<br>Public safety<br>Smart city | Consortium and solution platforms<br>Ecosystem orchestration<br>Multi-stakeholder delegation and collaboration<br>Industry alliances |
| **6. Mobile access provisioning** | Market penetration<br>Scale economics | Lock-in<br>Adaptability | Incumbent born platform<br>Smart grids | Solution enabling platform<br>Unified governance |

|  | Connection | SLA differentiation<br>Ubiquitous coverage<br>National roaming<br>Local sub-networks<br>Collaborative network<br>  sharing<br>Total defence dual-use | Industrial IoT<br>Cloud<br>Satellite operators<br>Public safety<br>Extended hospital<br>Assisted mobility | B2B complementors<br>Planning and controlling<br>Standardization |
|---|---|---|---|---|
| **7. Context awareness** | Market creation<br>Cross-side network<br>  effects<br>Content, context,<br>  commerce | Novelty<br>Awareness, immersive<br>  mapping<br>Data platform<br>  sovereignty<br>Real-time digital twin<br>Threat intelligence<br>Spectrum anomaly<br>  detection | Born platforms<br>Creating and connecting<br>  new market structures<br>Cloud<br>Sensor fusion and curation<br>Data AI federated learning | Transaction platform<br>Innovation and business<br>  ecosystems<br>Regulated governance<br>Ecosystem orchestration |
| **8. Service and performance** | Market creation<br>Cross-side network<br>  effects<br>Pay for success<br>User-oriented<br>  service<br>Outcome-based<br>  services<br>Commerce | Novelty<br>Adaptability<br>Cyber-physical<br>  integration<br>Network monetization<br>Application awareness<br>Sovereignty centric-<br>  model | Born platforms<br>Creating and connecting<br>  new market structures<br>Horizontal<br>Developer ecosystem<br>Pro-innovation, training | Application marketplace<br>  platforms<br>Regulated governance<br>Ecosystem orchestration<br>Co-evolution and<br>  orchestration |
| **9. Social mission, empowerment** | Market creation<br>Cross-side network<br>  effects<br>Sharing economy<br>Connection,<br>  computing,<br>  content, context,<br>  commerce | Complementarities<br>Adaptability, awareness<br>Resource sharing<br>Common resources<br>Data and context<br>  externalities<br>Trusted environment<br>Accessibility and<br>  inclusion<br>Community pooling | Born platforms<br>Smart cities, communities<br>Remote access inclusion<br>Developers<br>Ethical hackers<br>Prosumers<br>Knowledge brokers<br>NGOs, environmental<br>  reconstruction | Decentralised autonomous<br>  platform<br>Participatory/algorithmic<br>  governance<br>Co-evolution and<br>  orchestration<br>Open source<br>Frugal innovations |

### 6.4.1 Analysis and positioning of business model pattern groups

**Business model pattern**

Firms providing mobile access connectivity built on innovative technology design have traditionally adopted a market penetration business model pattern, aiming to achieve economies of scale. This model is built on tangible assets, standardisation, and centralised, vertically integrated functional structures for planning and controlling. The principal new business model that we see here is *market segmentation* expanding into new market segments, building on co-operative customisation and multi-divisional and matrix structure operations. The business model found emerging in a resilient 6G system is a *market creation* model that aids in the pursuit of opportunities generated by intersecting technologies and markets to develop and commercialise products for complementary markets. Effective market exploration, in context awareness, performance service, and social mission models, requires the development and use of community-based organisational designs and facilitative management approaches that enable firms in a particular area of economic activity to collaborate with their customers as well as with one another. For a firm to be successful in a complex and growing environment, such as a resilient 6G system, it is essential to have the capability to continually create, share, and apply knowledge to interact collaboratively within networks and communities (Miles, et al., 2009)

**Operating model (Modus)**

The *operating model approach (Modus)* of the identified business model groups and patterns was found to be structured around several distinct platform-based archetypes (Springer, et al., 2025) and their hybrids. *Consortium platforms* foster industrial collaboration by connecting public and private actors for shared value. Their pricing and co-operative models enable integration across industries through shared standards, facilitating large-scale interactions and network effects. They connect physical assets to build cyber-physical ecosystems, with data integration crucial for collaboration, optimisation, and decision-making in complex environments.

*Solution enabler platforms* found in co-operative and mobile access BM patterns integrate manufacturers, service providers, and end-users, emphasising co-innovation, tailored solutions, and trust-dependent transactions for highly specialised industries with smaller, niche user groups. While network effects are limited, the quality of the network and ecosystem is key. Integration with complementary platforms, domain technologies, and services supports tailored, value-driven modular solutions for specific industries. Extensive interoperability with physical assets such as sensing and IoT devices is essential for custom solutions and value creation. The platform's functionality depends on robust data sharing and governance frameworks suited to industrial needs.

*Transactional platforms* in context awareness connect user groups, lower search and transaction costs by enabling interactions and participation across industries and building trust among stakeholders. They have strong potential for cross-side network effects, mainly facilitating digital interactions and system integration. Data use is limited to improving recommendations, reducing costs, and enhancing user experience.

*Application marketplace platforms* connect industrial users with specialised developers and service providers, creating an ecosystem that encourages adoption and innovation within connected environments, and has cross-side network effects. The platform architecture supports transactions via standardised interfaces that promote interoperability and innovation, reducing multi-platform complexity. Data integration is key to the platform's role, enabling real-time operations through 6G network data, IoT devices, and analytics, enhancing productivity and efficiency.

*Decentralised autonomous platform*-based BMs identified in the digital supply chain and social mission serve the diversity of actors with distributed roles such as developers, users, and regulators, fostering collaborative innovation, trust, transparency, traceability, and cross-side network effects. Interoperability and scalability are achieved through open standards and decentralised protocols, with data integration complexity managed through distributed ledger technologies. Integration with physical assets enables seamless interaction and data exchange between digital platforms and physical assets across broad, decentralised networks and applications, ensuring secure and transparent tracking of asset ownership, usage, or transactions.

**Industrial clustering (Locus)**

From an industrial clustering perspective, the platform-based business model approach can be characterised by the focal actor. *Incumbent-born platforms* leverage patterns in pricing, design, supply chain, and mobile access. BM patterns are typically upstream platforms and are dependent on a core product. *Platform-born adjacent platforms* serve downstream users via transformative service innovations, particularly in co-operative patterns. The third model identified in co-operative, context awareness, performance service, and social patterns was a *born-platform* that is a stand-alone multi-sided platform type building on a digital platform value proposition from the beginning of a new venture, aiming at new market creation.

**Value creation (Focus)**

Complex networks, such as wireless and smart grids, involve dynamic, complex interactions between components that cannot be predicted by analysing parts alone. Additionally, natural and human-caused disruptive events cannot be fully prevented, underscoring the need for improved effectiveness of the system. Such a complex system design can be measured as the combined effect of reliability, robustness, and resilience (Zissis, 2019) as illustrated in Figure 6.3.

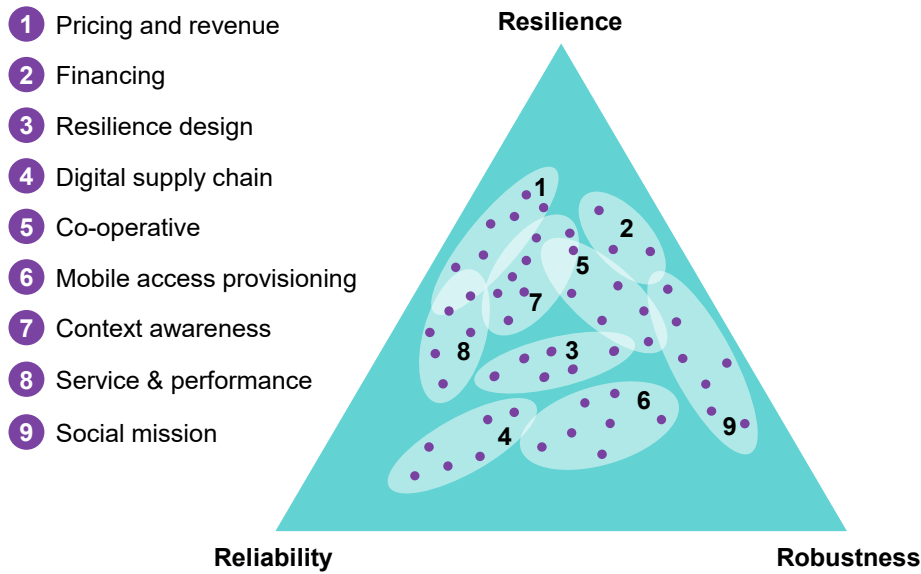

*Figure 6.3 The summary of the identified resilience business model pattern groups and value creation (Focus) patterns, along with their positioning in the 3R triangle.*

The leading pattern group contributing to *reliability* is within the digital supply chain (4), while mobile access provisioning (6) and social mission (9) support *robustness,* see **Error! Reference source not found.**. Service and performance patterns (8) leverage reliability and contribute to resilience via applications. Mobile communications primarily targeting *resilience* are not commonplace, though complementary pricing and revenue (1) and financing (2) were highlighted, particularly associated with public-private partnerships. The resilience design (3), co-operative (5), and context awareness (7) pattern groups were optimally positioned to have the opportunity to create value across all three 3R dimensions equally. Results indicate that standardisation, policy, and regulation should complement resilience innovations, given that the impacts of resilience have not yet been fully factored into economic systems.

## 6.5 Recommendations and future research

Envisioning a highly functional, dynamic, and sustainable 6G system as a GPT should target 1) *Resilience* as the ability of a system to recover from perturbation and to restore/ repair/ bounce back after a change due to an outside force, 2) *Self-organization* as the ability of a system to structure itself, to create new structure, to learn, diversify, complexify, and evolve, and 3) *Hierarchy* as the ability to evolve from the lowest level up to create a larger system via subsystems within systems (Meadows, D. H., 2008). At the same time, the identified architectural and technological enablers must focus on managing keystone *vulnerabilities* via protective design measures, developing *situational self-awareness* capabilities, and enhancing *adaptive* and reconfiguration capacity (McManus, et al., 2008) (Khaloopour, et al., 2024).

### 6.5.1 Business

From the business perspective, *service*-oriented paradigms transform 6G into *multi-layered platforms* and *ecosystems*. Open architecture and interfaces control of 6G networks within *applications* and novel *data-driven* business models will become the key factor in capturing value from AI-native 6G. This capability will have transformative implications for technology, regulation, and the economy. E.g., distributed cloud providers hosting both applications and networks, and TN–NTN system integration via MNO-SNO collaboration. New advantages can emerge from controlling the sovereignty of the strategic 6G infrastructure, leveraging legality by design and full regulatory compliance, considering the AI Act, Cyber Resilience Act (CRA), and Network Information System (NIS2) directive. This requires a combination of public-private partnership funding and a new financing model for innovations such as defence venture capital funding. Value appropriability potential and traditional mechanisms of intellectual property protection are shifting towards a more *transactional* and *collaborative* approach. This involves leveraging open disclosure and collaborative actions to foster the widespread adoption of innovations and knowledge. General-purpose 6G technology can be seen as a catalyst and contributor to resilience in other critical vertical sectors, such as utilities, public safety, healthcare, banking, logistics, and defence. Compared to consumer platforms, 6G business-to-business (B2B) platforms operate with smaller network sizes and weaker network effects, resulting in distinct market dynamics and complex power interdependencies, such as complementor bargaining power (Springer, et al., 2025).

*Platform governance and ownership* are crucial for the scalability of 6G. Analysing value creation, assets, and externalities is key to understanding the evolving 6G ecosystem with multiple platform layers. Value creation in 6G platforms is limited by standardisation, interoperability, system integration, and knowledge exchange challenges (Cusumano, et al., 2020). *Externalities* from innovation spillovers can cause system failures due to weak incentives for developing support components. Transaction platforms, which coordinate consumption, face issues like information asymmetry, matching, costs, and incentivising complementors. Consumption externalities may cause market failures from coordination problems hindering value exchange. Business ecosystems, supporting inter-firm cooperation, face 6G challenges like solution delivery, co-specialisation, free riding, bottlenecks, data, and interdependence. Co-operation externalities, where investments create ecosystem-wide value beyond individual gains, and bottlenecks in co-specialisation, can lead to value network failures due to low incentives for cooperation and joint value creation (Jacobides, et al., 2024).

Although platform and ecosystem models are efficient compared to traditional contracting and vertical integration by reducing externalities, they can also face significant *value architecture failures*, both functionally and distributionally (Jacobides, et al., 2024). Functional value architecture failures happen when joint value creation, the value proposition, and innovation don't align with resources and capabilities. In 6G platforms, this leads to issues like instability, shared interface problems, incompatible standards or applications, decentralised tech adoption causing fragmented products, and upgrade difficulties. Ecosystem failures occur when attracting complementors with an appealing value proposition fails, often due to poor design, competition, or uncoordinated decisions. Distributional failures arise when value capture doesn't match contributions, caused by imbalanced sharing among owners, complementors, and others. Ecosystems are especially prone to these issues due to the difficulty in balancing co-opetition (co-operation and competition) to keep actors aligned.

Extending a business model concept from a descriptive phenomenon to an explanatory and predictive theory opens new research avenues. Focusing only on BM design elements, research lacks enough empirical insight into the creation process of BMs, leaving open questions:

- How to ensure the openness of the 6G system for complementary and adjacent platform-based services and applications in different sectors?
- How to mitigate distinct types of risk that arise with ecosystems: co-innovation risks and adaptation chain risks? How can value contributions and value appropriation sharing be balanced among stakeholders, including the platform owner, complementors, and side actors?
- How can the real-time data orchestration (collect, store, sort, analyse, interpret) capability and utilisation with AI-based digital twinning become a competitive advantage through innovating future superior offerings as a part of the product development process?

- How to balance Business to Business (B2B) and Business to Government (B2G) regulatory and business processes, such as export control, tendering process, security clearances, and alliances?
- From an organisational perspective, how do enterprises organise themselves beyond the focal enterprise to implement business model innovation, and how are the design and implementation processes planned and executed?

### 6.5.2 Regulation

From a regulatory perspective, resilience has become a key goal in telecommunications. The European Commission addresses the connected, collaborative computing (3C) network as a European telco edge cloud platform (European Commission, 2024) and technology aggregator (Draghi, M., 2024). Communication networks exhibit a dualistic nature, serving as both a catalyst for social cohesion and collective action, while simultaneously harbouring the potential to exacerbate societal divisions. Insufficient regulation exposes individuals, the environment, and democratic processes to risk, while simultaneously harming the business environment through increased administrative burdens and legal uncertainty. This can erode public trust, impede the responsible adoption of technology, and leave markets and individuals vulnerable to inefficiencies. A regulatory void may lead to a deficiency in existing legal and policy frameworks, creating an area of unchecked activity with potential negative consequences for sustainability, resilience, and economic equity. Regulatory loopholes, on the other hand, represent flaws in legal frameworks that enable the circumvention of intended regulatory objectives, often within the bounds of the law.

The European regulatory environment is characterised by complexity and a lack of regulatory flexibility, making it difficult to predict the cumulative impact of existing legislation. The regulatory landscape should increasingly be characterised by *experimental* frameworks and *agile* governance mechanisms, including collaborative regulatory approaches to enhance policy coherence, facilitate experimentation, and allow for more responsive and adaptable regulation. Regulatory sandboxes provide controlled environments for testing innovative products and services under regulatory oversight. This facilitates the regulator's understanding of emerging technologies and market dynamics, enabling the market entry of beneficial innovations while maintaining safety standards. The resulting dialogue between regulators and innovators enables the development of adaptive regulatory frameworks that strike a balance between innovation and resilience, thereby promoting societal acceptance of new technologies.

The intervention of public and technology policy and regulation is crucial to maintaining realistic expectations. A holistic policy framework that addresses resilience across the entire telecom ecosystem involves collaboration between governments, regulators, and industry players, covering interoperability from organisational to platform and technical layers. Without this complementary approach, it would be unrealistic to expect novel technological enablers to deliver the desired resilience outcomes. The expectations of resilience also introduce new policy and regulatory frameworks.

*Market balance* can be improved by "same rules for same services," ensuring fair commercial outcomes in the ecosystem by addressing the dominance of large hyperscalers and online service providers and ensuring sustainable investments in network infrastructures. This involves creating a *level playing field* that allows new business models to emerge and fosters innovation within the digital market.

Short-term *national* regulatory strategies are required, such as temporary disaster roaming (TDR), resource reuse and sharing, and incentives for operator collaboration towards borderless networks. The digital product passport as an SBOM can assist in resilient digital supply chain design for products, applications, and services. Ambitious national broadband strategies often set targets for universal access but lack clear implementation paths. Cross-border services involving the management of personal data will need to be regulated by a diverse set of national rules and entities.

A resilient 6G system will never be optimal with respect to traditional performance indicators related to efficiency due to the required surplus resources. Hybrid control architectures loosely coupled subsystems, and virtualised elements may lead to performance losses and impact security procedures, e.g., via zero trust architecture overhead and trade-offs. Longer-term research action is needed towards a fully resilient-by-design 6G system that co-optimises resilience-sustainability-economic trade-offs, particularly related to complexity/cost, end-to-end usage of resources, and reconfigurability.

### 6.5.3 Standardization

Resilience has gradually entered the standardisation domain. The United Nations' International Telecommunication Union Radiocommunication Sector (ITU-R) has introduced four overarching design principles as new elements to guide the global 6G system development: sustainability, connecting the unconnected, ubiquitous intelligence, and security and resilience (ITU-R, 2023). Resilience "*refers to capabilities of the networks and systems to continue operating correctly during and after a natural or man-made disturbance, such as the loss of primary source of power, etc*" (ITU-R, 2023). ITU-T Recommendation G.827 (ITU-T, 2023) defines network performance parameters and objectives for the path elements and end-to-end availability of international constant bit-rate digital paths. These parameters are independent of the underlying physical network technology (e.g., optical fibre, mobile, or satellite). The recommendations guide improvements in availability and calculating end-to-end availability for combinations of network elements. Implicitly, it promotes network design principles that enhance resilience through redundancy, diverse routing, fast protection switching, and a focus on end-to-end performance. The document offers a framework for quantifying and improving availability, a direct indicator of network resilience to failures.

The 3GPP study on the 6G use cases and service requirements (3GPP, 2025) addresses potential new requirements for network resiliency related to use cases such as quantum-resistant security, zero-outage network, fast network provisioning to improve resilience, ubiquitous and resilient TN and NTN, resilient positioning in satellite networks, and utility infrastructure monitor and control.

Seamless end-to-end management across domains governed by different standards (e.g., 3GPP vs IEEE) remains a significant research challenge. Medium-term cross-standardisation and application actions between mobile communications and IT in 3GPP, IEEE, ETSI, O-RAN, NIST, and ISO/IEC, and with various verticals, are needed to cope with cybersecurity and enable different operational modes in the resilience context. Recently, the newly widened positioning of ETSI in IT standardisation has been discussed, for example.

# Abbreviations

| | | | |
|---|---|---|---|
| **3D** | 3-Dimensional | **ML** | Machine Learning |
| **3GPP** | 3rd Generation Partnership Project | **MNO** | Mobile Network Operator |
| **6G** | Sixth Generation | **MRSS** | Multi-Radio Spectrum Sharing |
| **AI** | Artificial Intelligence | **MTD** | Moving Target Defence |
| **AI-RAN** | AI-Optimised Radio Access Networks | **NAS** | National Academy Of Sciences |
| **API** | Application Programming Interface | **NFV** | Network Functions Virtualisation |
| **B2G** | Business To Government | **NIS** | Network Information System |
| **BM** | The Business Model | **NIST** | National Institute Of Standards And Technology |
| **BS** | Base Station | **NTN** | Non-Terrestrial Network |
| **CAPEX** | Capital Expenditure | **NTP** | Network Time Protocol |
| **CERRE** | Centre On Regulation in Europe | **O-RAN** | Open RAN |
| **CNI** | Container Network Interface | **OPEX** | Operating Expense |
| **CR** | Cognitive Radio | **PHY** | Physical Layer |
| **CRA** | Cyber Resilience Act | **PKI** | Public Key Infrastructure |
| **CRMA** | Critical Raw Materials Act | **PPP** | Public-Private Partnership |
| **CTC** | Cross-Technology Communication | **PQC** | Post-Quantum Cryptography |
| **CU** | Central Unit | **PTP** | Precision Time Protocol |
| **D-MIMO** | Distributed Massive MIMO | **QOS** | Quality Of Service |
| **D2D** | Direct-To Device | **RACH** | Random Access Channel |
| **DER** | Distributed Energy Resources | **RAG** | Retrieval-Augmented Generation |
| **DLP** | Discrete Logarithmic Problem | **RAN** | Radio Access Network |
| **DNA** | Digital Networks Act | **RAT** | Radio Access Technology |
| **DTC** | The Direct-To-Cell | **RF** | Radio Frequency |
| **DU** | Distributed Unit | **RIC** | Ran Intelligent Controller |
| **E2E** | End-To-End | **RIS** | Reconfigurable Intelligent Surfaces |
| **EMBB** | Enhanced Mobile Broadband | **RRC** | Radio Resource Control |
| **ETSI** | European Telecommunications Standards Institute | **RT** | Real Time |
| **EU** | European Union | **RU** | Radio Unit |
| **FRMCS** | Future Railway Mobile Communication System | **SDN** | Software-Defined Networking |
| **GANA** | Generic Autonomic Networking Architecture | **SFA** | Strategic Foresight Analysis |
| **GEO** | Geostationary Orbit | **SL** | D2D Sidelink |
| **GNSS** | Global Navigation Satellite Systems | **SLA** | Service-Level Agreement |
| **HAPS** | High-Altitude Platform Station | **SW** | Software |
| **IOT** | Internet Of Things | **TDR** | Temporary Disaster Roaming |
| **IRGC** | EPFL International Risk Governance Centre | **TLS** | Transport Layer Security |
| **ISL** | Inter-Satellite Link | **TN** | Terrestrial Network |
| **ITU** | International Telecommunications Union | **TSN** | Time-Sensitive Networking |
| **ITU-R** | ITU-Radiocommunication Sector | **UAV** | Unmanned Aerial Vehicle |
| **KPI** | Key Performance Indicator | **UE** | User Equipment |
| **LEO** | Low Earth Orbit | **URLLC** | Ultra-Reliable Low-Latency Communication |
| **LPWAN** | Low-Power WAN | **VLAN** | Virtual Local Area Network |
| **MAC** | Media Access Control Layer | **VSAT** | Very Small Aperture Terminal |
| **MEC** | Multi-Access Edge Computing | **WAB** | Wireless Access Backhaul |
| **MEO** | Medium Earth Orbit | **WAN** | Wide-Area Network |
| **MIMO** | Multiple Input Multiple Output | | |

# References

3GPP, 2025. *Chair's summary of the 3GPP workshop on 6G (6GWS-250243),* s.l.: 3GPP.

3GPP, 2025. *TR 22.870 V0.2.1 Study on 6G Use Cases and Service Requirements, Stage 1, Release 20.,* s.l.: 3GPP.

Aalto University, 2013. *Cyclone Dagmar's impact on Finland's power grid,* Helsinki: Aalto University.

Ahmadian, N., Lim, G. J., Cho, J. & Bora, S., 2020. A quantitative approach for assessment and improvement of network resilience. *Reliability Engineering & System Safety,* August, ( ), p. .

Ahokangas, P. ym., 2023. Toward an integrated framework for developing European 6G innovation.. *Telecommunications Policy,* 47(9), p. 102641.

Ahokangas, P., Matinmikko-Blue, M. & Yrjölä, S., 2022. Envisioning a future-proof global 6G from business, regulation, and technology perspectives.. *IEEE Communications Magazine,* 61(2), pp. 72-78.

Ale, L., King, S. A., Zhang, N. & Xing, H., 2025. Enhancing generative AI reliability via agentic AI in 6G-enabled edge computing. *Nature Reviews Electrical Engineering,* p. 1–3.

Alves, H., Mikhaylov, K. & Höyhtyä, M., 2024. *Integration of MTC and Satellites for IoT toward 6G era.* 1 ed. Hoboken: Wiley/IEEE Press.

Amit, R. & Zott, C., 2001. Value creation in e-business.. *Strategic Management Journal,* 22(6-7), pp. 493-520.

Andersson, C. et al., 2022. *Improving energy performance in 5G networks and beyond,* : Ericsson Technology Review.

Bennis, M., 2025. *Resilient-native and Intelligent NextG Systems.* : .

Benzaïd, C. ym., 2025. *A Multi-Layered Zero Trust Microsegmentation Solution for Cloud-Native 5G & Beyond Networks.* s.l., s.n., p. 1–7.

Bertsekas, D., 2021. *Rollout, policy iteration, and distributed reinforcement learning.* s.l.:Athena Scientific.

Bertsekas, D., 2022. *Abstract dynamic programming.* s.l.:Athena Scientific.

Buldyrev, S. V. et al., 2010. Catastrophic cascade of failures in interdependent networks. *Nature,* Volume 464, p. 1025–1028.

Chesbrough, H. & Rosenbloom, R. S., 2002. The role of the business model in capturing value from innovation: evidence from Xerox Corporation's technology spin-off companies.. *Industrial and Corporate Change,* 11(3), pp. 529-555.

Curic, M. & van Maastricht, C., 2024. *Applying generative AI to revolutionize telco network operations.* s.l.:s.n.

Cusumano, M., Yoffie, D. & Gawer, A., 2020. *The future of platforms..* Cambridge, MA: MIT Sloan Management Review.

Department of Energy, 2017. *Small Modular Reactors: Adding to Resilience at Federal Facilities,* Idaho: U.S. Department of Energy.

Draghi, M., 2024. *The future of European competitiveness: A competitiveness strategy for Europe.,* s.l.: European Commission.

Elkington, J. & Rowlands, I. H., 1999. Cannibals with forks: The triple bottom line of 21st century business.. *Alternatives Journal,* 25(4), p. 42.

Ergenç, D. ym., 2025. Resilience in Edge Computing: Challenges and Concepts. *Foundations and Trends® in Networking,* Osa/vuosikerta 14, p. 254–340.

Ericsson, 2025. *Global and regional key figures – Ericsson Mobility Report,* s.l.: Ericsson.

European Commission, 2023. *An EU approach to enhance economic security.,* s.l.: European Commission.

European Commission, 2024. *WHITE PAPER How to master Europe's digital infrastructure needs?,* s.l.: European Commission.

European Union, 2024. *EU 2024/1252 of the European parliament and of the council: Establishing a framework for ensuring a secure and sustainable supply of critical raw materials and amending Regulations (EU) No 168/2013, (EU) 2018/858, (EU) 2018/1724 and (EU) 2019/1020.,* s.l.: European Union.

Financial Times, 2024. *Chile reels after worst blackout in 15 years hits copper mines.* [Online] Available at: https://www.ft.com/content/25f681c9-a3dd-4af5-b065-a2d15e14fe88 [Accessed 2024].

Foss, N. J. & Saebi, T., 2017. Fifteen years of research on business model innovation: How far have we come, and where should we go?. *Journal of Management,* 43(1), pp. 200-227.

Gholipoor, N., Aghdam, F. H., Rasti, M. & Zarini, H., 2025. Strategic Utilization of Cellular Operator Energy Storage for Smart Grid Frequency Regulation. *IEEE Transactions on Smart Grid.*

GSMA, 2024. *The Mobile Economy 2024,* s.l.: GSMA.

Hallegatte, S., Rentschler, J. & Rozenberg, J., 2019. *The Resilient Infrastructure Opportunity,* : World Bank.

Heinimann, H. R., 2016. A Generic Framework for Resilience Assessment. In: M. Florin & I. Linkov, eds. *IRGC resource guide on resilience*. Lusanne: EPFL, pp. 90-95.

Heinimann, H. R., 2018. Robustness and Reconfigurability – Key Concepts to Build Resilience. In: B. Trump, M. Florin & I. Linkov, eds. *IRGC resource guide on resilience (vol. 2): Domains of resilience for complex interconnected systems*. Lusanne: EPFL International Risk Governance Center (IRGC).

Hewa, T., Braeken, A., Liyanage, M. & Ylianttila, M., 2022. Fog computing and blockchain-based security service architecture for 5G industrial IoT-enabled cloud manufacturing. *IEEE transactions on industrial informatics,* Osa/vuosikerta 18, p. 7174–7185.

Hutchison, D., Pezaros, D., Rak, J. & Smith, P., 2023. On the Importance of Resilience Engineering for Networked Systems in a Changing World. *IEEE Communications Magazine,* November, 61(11), pp. 200-206.

ITU, 2023. *Framework and overall objectives of the future development of IMT for 2030 and beyond,* s.l.: ITU.

ITU-R, 2023. *Recommendation M.2160-0. Framework and overall objectives of the future development of IMT for 2030 and beyond.,* s.l.: ITU-R.

ITU-T, 2023. *Recommendation G.827. Digital networks – Quality and availability targets. Series G: Transmission systems and media, digital systems and networks.,* s.l.: ITU-T.

Jacobides, M. G., Cennamo, C. & Gawer, A., 2024. Externalities and complementarities in platforms and ecosystems: From structural solutions to endogenous failures.. *Research Policy,* 53(1), p. 104906.

Javadpour, A., Ja'fari, F., Taleb, T. & Benzaïd, C., 2024. *5G Slice Mutation to Overcome Distributed Denial of Service Attacks Using Reinforcement Learning.* s.l., s.n., p. 1–9.

Karacora, Y., Chaccour, C., Sezgin, A. & Saad, W., 2024. *Event-Based Framework for Agile Resilience in Criticality-Aware Wireless Networks*. : .

Katz, M. & Perera, A., 2023. Novel data and energy networking in WPAN networks. *IEEE Wireless Communications*.

Kemene, E. & Christianson, A., 2025. *What we can learn about building a resilient energy grid from the Iberian power outage*. [Online]
Available at: https://www.weforum.org/stories/2025/05/resilient-energy-grid-iberian-power-outage/
[Accessed 7 Aug. 2025].

Khaloopour, L. et al., 2024. Resilience-by-Design in 6G Networks: Literature Review and Novel Enabling Concepts.. *IEEE Access,* 12(-), pp. 155666-155695.

Kianpisheh, S., Benzaïd, C. & Taleb, T., 2024. *Multi-model based federated learning against model poisoning attack: A deep learning based model selection for MEC systems*. s.l., s.n., p. 1737–1742.

Koç, Y. et al., 2014. The impact of the topology on cascading failures in a power grid model.. *Physica A: Statistical Mechanics and its Applications,* 402(-), pp. 169-179.

Laprie, J., 1992. Dependability: Basic Concepts and Terminology. Teoksessa: J. Laprie, toim. *Dependability: Basic Concepts and Terminology. Dependable Computing and Fault-Tolerant Systems.* Vienna: Springer, p. .

Lee, H., Kim, S. & Kim, H. K., 2022. *SoK: Demystifying Cyber Resilience Quantification in Cyber-Physical Systems*. Rhodes, Greece, IEEE, pp. 178-183.

Lüdeke-Freund, F. ym., 2018. The sustainable business model pattern taxonomy—45 patterns to support sustainability-oriented business model innovation.. *Sustainable Production and Consumption,* 15(-), pp. 145-162.

Mahmood, N. H. et al., 2025. Resilient-By-Design: A Resiliency Framework for Future Wireless Networks. *IEEE Communications Magazine,* ( ), p. .

Manzano, M. et al., 2014. Robustness surfaces of complex networks. *Scientific Reports,* 4(6133).

Matinmikko-Blue, M. et al., 2020. *White paper on 6G drivers and the UN SDGs.,* s.l.: arXiv.

McManus, S., Seville, E., Vargo, J. & Brunsdon, D., 2008. A facilitated process for improving organizational resilience.. *Natural Hazards Review,* 9(-), p. 81–90.

Meadows, D. H., 2008. *Thinking in systems: A primer.*. s.l.:Chelsea Green Publishing.

Mieghem, P. V. et al., 2010. *A Framework for Computing Topological Network Robustness,* : Delft University of Technology.

Miles, R. E. ym., 2009. The I-Form Organization.. *California Management Review,* 51(4), pp. 61-76.

Motalleb, M. K. ym., 2025. Towards secure intelligent O-RAN architecture: vulnerabilities, threats and promising technical solutions using LLMs. *Digital Communications and Networks*.

NATO, 2023. *Allied command transformation, Strategic Foresight Analysis (SFA).,* s.l.: NATO.

Nguyen, N. T., Lee, K. & DaiIEEE, H., 2021. Application of deep learning to sphere decoding for large MIMO systems. *IEEE Transactions on Wireless Communications,* Osa/vuosikerta 20, p. 6787–6803.

Onetti, A., Zucchella, A., Jones, M. V. & McDougall-Covin, P. P., 2012. Internationalization, innovation and entrepreneurship: business models for new technology-based firms.. *Journal of Management & Governance,* 16(-), pp. 337-368.

Perera, A., Godaliyadda, R. & J. Hakkinen, M. K., 2025. Lighting the way for a sustainable future: Overcoming challenges in light-based IoT and data-energy networking. *IEEE Communications Magazine,* pp. 1-7.

Rak, J. & Hutchison, D. eds., 2020. *Guide to Disaster-Resilient Communication Networks*. Cham: Springer.

Rak, J. & Hutchison, D., 2020. *Guide to Disaster-Resilient Communication Networks..* 1 ed. Cham, Switzerland: Springer.

Reifert, R.-J.et al., 2024. Resilience and Criticality: Brothers in Arms for 6G. *arXiv,* ( ), p. .

Reifert, R. -J., Roth, S., Ahmad, A. A. & Sezgin, A., 2023. Comeback Kid: Resilience for Mixed-Critical Wireless Network Resource Management. *IEEE Transactions on Vehicular Technology,* December, 72(12), pp. 16177-16194.

Rezaki, A. et al., 2025. *Sustainability in SNS JU Projects - Targets, Methodologies, Trade-offs and Implementation Considerations Towards 6G Systems,* : .

Saarnisaari, H., Markkula, J., Paso, T. & Bräysy, T., 2021. Internetwork Time Synchronization of Mobile Ad Hoc Networks. *IEEE Access,* Volume 9, pp. 84191-84203.

Sarliève, P. et al., 2025. *A mapping tool for digital regulatory frameworks: Including a pilot on efforts to regulate AI.,* Paris: OECD Publishing.

Shafi, M. et al., 2017. 5G: A tutorial overview of standards, trials, challenges, deployment, and practice. *IEEE Journal on Selected Areas in Communications,* 35(6), pp. 1201-1221.

Shalev-Shwartz, S. & others, 2012. Online learning and online convex optimization. *Foundations and Trends® in Machine Learning,* Osa/vuosikerta 4, p. 107–194.

Shanmugam, K. ym., 2013. Femtocaching: Wireless content delivery through distributed caching helpers. *IEEE transactions on information theory,* Osa/vuosikerta 59, p. 8402–8413.

Sharma, N., Tabandeh, A. & Gardoni, P., 2018. Resilience analysis: a mathematical formulation to. *Sustainable and Resilient Infrastructure,* 3(2), pp. 49-67.

Shor, P. W., 1994. *Algorithms for quantum computation: discrete logarithms and factoring.* s.l., s.n., p. 124–134.

Silver, D. ym., 2017. Mastering the game of Go without human knowledge. *nature,* Osa/vuosikerta 550, p. 354–359.

Siriwardhana, Y. ym., 2024. Shield-secure aggregation against poisoning in hierarchical federated learning. *IEEE Transactions on Dependable and Secure Computing*.

SNS JU, 2025. *White paper 6G KVIs – SNS projects initial survey results 2025.,* s.l.: Smart Networks and Services Joint Undertaking (SNS JU).

Spence, M., 2011. Strategic Capitalism: A New Model for the 21st Century.. *Foreign Affairs,* 90(3), pp. 2-11.

Springer, V. ym., 2025. Platform design and governance in industrial markets: Charting the meta-organizational logic.. *Research Policy,* 54(6), p. 105236.

Srnicek, N., 2017. *Platform capitalism..* s.l.:John Wiley & Sons.

Sterbenz, J. et al., 2010. Resilience and Survivability in Communication Networks: Strategies, Principles, and Survey of Disciplines. *Computer Networks,* 54(8), pp. 1245-1265.

Stockholm Resilience Centre, 2020. *Resilience dictionary,* s.l.: Stockholm Resilience Centre.

Teece, D. J., 2018. Profiting from innovation in the digital economy: Enabling technologies, standards, and licensing models in the wireless world.. *Research Policy,* 47(8), pp. 1367-1387.

Timmers, P., 2024. *Resilience in Digital Supply Chains: Opportunities for Global and International Governance.,* s.l.: Centre on Regulation in Europe (CERRE).

Ungar, M., ed., 2021. *Multisystemic Resilience: Adaptation and Transformation in Contexts of Change.* ed. : Oxford University Press.

Vespignani, A., 2010. The fragility of interdependency. *Nature,* Volume 464, pp. 984-985.

Weick, K. E., Sutcliffe, K. M. & Obstfeld, D., 1999. Organizing for high reliability: Processes of collective mindfulness. *Research in organizational behavior,* 21( ), pp. 81-123.

Wirtz, B. W., Schilke, O. & Ullrich, S., 2010. Strategic development of business models: implications of the Web 2.0 for creating value on the internet.. *Long Range Planning,* 43(2), pp. 272-290.

Yrjölä, S., 2024. *Business models and profiting from innovation in future mobile communications.,* s.l.: Acta Universitatis Ouluensis.

Yrjölä, S., Matinmikko-Blue, M. & Ahokangas, P., 2023. *Developing 6G visions with stakeholder analysis of 6G ecosystem..* -, IEEE, pp. 705-710.

Zappone, A., Di Renzo, M. & Debbah, M., 2019. Wireless networks design in the era of deep learning: Model-based, AI-based, or both?. *IEEE Transactions on Communications,* Osa/vuosikerta 67, p. 7331–7376.

Zhu, Z., Hong, J., Drew, S. & Zhou, J., 2022. Resilient and communication efficient learning for heterogeneous federated systems. *Proceedings of machine learning research,* Osa/vuosikerta 162, p. 27504.

Zissis, G., 2019. The R3 Concept: Reliability, Robustness, and Resilience. *IEEE Industry Applications Magazine,* 25(4), pp. 5-6.

Zott, C. & Amit, R., 2010. Business model design: An activity system perspective.. *Long Range Planning,* 43(2-3), pp. 216-226.

Zubow, A. ym., 2025. *Towards Resilient and Efficient Multi-RAT Operation through Network Coding.* s.l., s.n., p. 1–4.

Zubow, A., von Stebut, I., Rösler, S. & Dressler, F., 2024. *ResCTC: Resilience in Wireless Networks through Cross-Technology Communication.* s.l., s.n., p. 1–6.